\documentclass[usenatbib]{mnras}

\usepackage{newtxtext,newtxmath}
\usepackage[T1]{fontenc}

\DeclareRobustCommand{\VAN}[3]{#2}
\let\VANthebibliography\thebibliography
\def\thebibliography{\DeclareRobustCommand{\VAN}[3]{##3}\VANthebibliography}

\usepackage{graphicx}	
\usepackage{amsmath}	
\usepackage{float}
\usepackage{soul}

\usepackage{xspace}
\usepackage[dvipsnames]{xcolor}
\newcommand{\azul}[1]{{{\color{blue} #1}}\xspace}

\title[Nitrogen abundance estimations in LINER galaxies]{Chemical abundances of LINER galaxies - Nitrogen abundance estimations}

\author[C.B. Oliveira Jr. et al.]{C. B. Oliveira Jr.$^{1}$\thanks{E-mail: cbo\_jr@hotmail.com (CBO)},
A.C. Krabbe$^{2}$\thanks{E-mail: angela.krabbe@gmail.com (ACK)},
O. L. Dors Jr.$^{1}$,
I. A. Zinchenko$^{3,4}$,
\newauthor{J. A. Hernandez-Jimenez$^{1}$,
M.~V. Cardaci$^{5}$,
G.~F. H\"agele$^{5}$,
G.~S. Ilha$^{2}$}\\
$^{1}$ Universidade do Vale do Para\'{\i}ba, Av. Shishima Hifumi, 2911, Zip Code 12244-000, S\~ao Jos\'e dos Campos, SP, Brazil\\
$^{2}$ Universidade de São Paulo, Instituto de Astronomia,
Geofísica e Ciências Atmosféricas, Rua do Matão 1226, CEP 05508-090, São Paulo, SP, Brazil\\
$^3$ Faculty of Physics, Ludwig-Maximilians-Universit\"{a}t, Scheinerstr. 1, 81679 Munich, Germany\\
$^4$ Main Astronomical Observatory, National Academy of Sciences of Ukraine, 27 Akad. Zabolotnoho St 03680 Kyiv, Ukraine\\
$^{5}$ Instituto de Astrof\'isica de La Plata (CONICET--UNLP), Paseo del Bosque s/n, B1900FWA La Plata, Argentina.}
\date{Accepted XXX. Received YYY; in original form ZZZ}

\pubyear{2023}

\begin{document}
\label{firstpage}
\pagerange{\pageref{firstpage}--\pageref{lastpage}}
\maketitle
\begin{abstract}
In this work, we investigated the nitrogen and oxygen abundances in a sample of galaxies with Low Ionization Nuclear Emission Regions (LINERs) in their nucleus. Optical spectroscopic data 
(3\,600 - 10\,000 \AA) of 40 LINERs from the Mapping Nearby Galaxies (MaNGA) survey were considered. Only objects classified as
retired galaxies, i.e. whose main ionization sources are post-Asymptotic Giant Branch (pAGB) stars, were selected.
The abundance estimates were obtained through detailed photoionization models built with the {\sc cloudy} code to reproduce a set of observational emission line intensities ratios of the sample. Our results show that LINERs have oxygen and 
nitrogen abundances in the ranges of 
$\rm 8.0 \: \la \: 12+\log(O/H) \: \la \: 9.0$
(mean value $8.74\pm 0.27$) and
$\rm 7.6 \: \la \: 12+\log(N/H) \: \la \: 8.5$  
(mean value $8.05\pm 0.25$), respectively. 
About 70\% of the sample have oversolar O/H and N/H
abundances. Our abundance estimates are in consonance with those for Seyfert~2 nuclei and \ion{H}{ii} regions with the highest metallicity, indicating that these distinct object classes show similar enrichment of the interstellar medium (ISM). 
The LINERs in our sample are located in the higher N/O region of the N/O versus O/H diagram, showing  an unexpected negative correlation between these two parameters. These results suggest that these LINERs mainly exhibit a secondary nitrogen production and could be acting some other mechanisms that deviate them from the usual theoretical secondary nitrogen production curve and the \ion{H}{ii} regions observations. However, we did not find any evidence in our data able to support the literature suggested mechanisms. On the other hand,
our results show that LINERs do not present any correlation between the N/O abundances and the stellar masses of the hosting galaxies.

\end{abstract}



\section{Introduction}

Estimations of chemical abundances of the gas phase of Active Galactic Nuclei (AGNs) and star-forming (SF) regions (\ion{H}{ii} regions and \ion{H}{ii} galaxies) are fundamental to understanding the chemical evolution of galaxies along the history of the Universe. The spectra of these objects present strong emission lines that are easily detectable, even in objects at (very) high redshift ($z $ = $\sim1.0$ - $\sim9.0$) (e.g., \citealt{2023MNRAS.518..425C, 2023arXiv230308149S}), and these lines can be used to estimate the metallicity ($Z$) and other properties (e.g., electron density, elemental abundance, hardness of the ionizing spectra, etc)  of these objects (for a review see \citealt{2019A&ARv..27....3M, 2019ARA&A..57..511K}). 

Generally, the gas-phase metallicity is traced through the oxygen abundance in relation to the hydrogen (O/H), due to the oxygen has strong emission lines in the optical spectrum ([\ion{O}{ii}]$\lambda3726,\lambda3729$, [\ion{O}{iii}]$\lambda5007$) emitted by its more abundant ions ($\rm O^{+}, O^{2+}$). Therefore, hereafter we use metallicity ($Z$) and
oxygen abundance (O/H) interchangeably. 
It is largely accepted that the most reliable method to estimate O/H and the abundance of other elements
(e.g., N, S, Ne) is the $T_{\rm e}$-method \cite[see discussion in][]{2006MNRAS.372..293H,2008MNRAS.383..209H}\footnote{For a review of the
$T_{\rm e}$-method see \citet{2017PASP..129h2001P} and \citet{2017PASP..129d3001P}.}.
The reliability of the $T_{\rm e}$-method  is supported by the agreement
between O/H estimates in \ion{H}{ii} regions and those in neighborhood
stars  (see \citealt{2003A&A...399.1003P, 2017MNRAS.467.3759T}).
The $T_{\rm e}$-method is based on determinations of electron temperatures and requires measurements of auroral emission lines, such [\ion{O}{iii}]$\lambda$4363 and [\ion{N}{ii}]$\lambda$5755, which are weak
(about 100 times weaker than H$\beta$)
or even not detectable in most parts of objects with high metallicity and/or low excitation (\citealt{1998AJ....116.2805V,2007MNRAS.382..251D, 2008A&A...482...59D}).  
In the cases in which the $T_{\rm e}$-method can not be applied, $Z$ and elemental abundances
(e.g., N/H, S/H) have been  estimated through  calibrations between these and 
intensities of strong emission line ratios (hereafter strong-line method) of SF regions (e.g., \citealt{1979MNRAS.189...95P, 1979A&A....78..200A, 1991ApJ...380..140M, 1996MNRAS.283..990T, 1997A&A...322...41C, 2000MNRAS.312..130D, 2000ApJ...539..687O, 2002ApJS..142...35K, 2004MNRAS.348L..59P,  2006A&A...449..193P, 2006A&A...454L.127S, 2006A&A...459...85N, 2007A&A...462..535Y, 2007MNRAS.381.1719V, 2007A&A...475..409S, 2013ApJS..208...10D, 2013A&A...559A.114M, 2014ApJ...797...81M, 2015ApJ...813..126J, 2016MNRAS.458.1529B, 2016MNRAS.457.3678P, 2017MNRAS.465.1384C, 2018AJ....155...82H, 2019ApJ...872..145J, 2019MNRAS.485.3569H, 2020A&A...636A..42M, 2021MNRAS.504.1237P, 2022MNRAS.513.2006F, 2022MNRAS.511.4377D}) and AGNs (\citealt{Storchi_Bergmann_1998, 10.1093/mnras/stx150, 2020MNRAS.492.5675C, 2021MNRAS.507..466D, 2021MNRAS.501.1370D}).

The vast spectroscopic optical data obtained by surveys -- e.g., data made available by
the Sloan Digital Sky Survey (SDSS, \citealt{2000AJ....120.1579Y}), Calar Alto Legacy Integral Field Area  (CALIFA, \citealt{2012A&A...538A...8S}),  Mapping Nearby Galaxies
at Apache Point Observatory (MaNGA, \citealt{2015ApJ...798....7B}), and Chemical Abundances of Spirals (CHAOS, \citealt{2015ApJ...806...16B})  -- have revolutionized 
the understanding of the chemical enrichment of SF regions (e.g., \citealt{2004ApJ...613..898T, 2006A&A...448..955I, 2019A&A...623A..40I, 2014A&A...563A..49S, 2017MNRAS.469..151B, 2020ApJ...893...96B, 2022ApJ...939...44R}, among others)  in the local universe.
However,  estimations of the metallicity and elemental abundances in other object classes, such as  AGNs and  Low Ionization Nuclear Emission-line Region (LINERs) are little known. For instance, the first quantitative determination of the N/H abundance in AGNs (Seyfert~2 type) was only performed recently by \citet{2017MNRAS.468L.113D}. These authors used detailed photoionization models to reproduce the optical spectrum of 44 Seyfert~2s in the local ($z\: < \: 0.1$) universe (see also \citealt{2019MNRAS.489.2652P, 2020MNRAS.496.2191F}). Moreover,
neon, argon, helium, and sulfur abundances have also been recently estimated in a small sample
(less than 70 objects, $z\: < \: 0.3$) of AGNs by 
\citet{2021MNRAS.508..371A}, \citet{2021MNRAS.508.3023M} and \citet{2022MNRAS.514.5506D, 2023MNRAS.521.1969D}, respectively.

A worse scenario is found for LINERs,  
despite these objects appearing in $\sim1/3$ of
galaxies in the local universe \citep{netzer_2013}.
This fact is possibly due to the need to know the nature of the ionizing sources and excitation mechanisms of LINERs 
in order to apply the $T_{\rm e}$-method \citep{2020MNRAS.496.3209D} and/or strong-line methods \citep{Storchi_Bergmann_1998}.
\citet{2010A&A...519A..40A}, using optical spectra of a sample of
LINERs located in early-type galaxies (ETGs), estimated the O/H abundance 
assuming hot stars and the accretion of gas into a central black hole (AGN) as ionization sources. The authors found that the average abundance obtained from AGN calibration is $\sim0.05$ dex higher than those obtained through hot stars calibration, however, the last one produced a broader metallicity range.   
\citet{2021MNRAS.tmp.1341K} derived the O/H abundance of the
 UGC\,4805 LINER nucleus using  MaNGA data and using  distinct methods, i.e.,   the extrapolation of the disk abundance gradient, the calibrations
between O/H abundance and strong emission lines for AGNs,  as well
as  photoionization models built with the \textsc{cloudy} code \citep{2017RMxAA..53..385F}, assuming gas accretion into a
black hole (AGN) and post-asymptotic giant branch (pAGB) stars with different effective
temperatures. 
 These authors found that depending on the method adopted, discrepancies of until 
 $\sim 0.4$ dex are derived
 (see Table~2 of \citealt{2021MNRAS.tmp.1341K}).
Finally, \citet{2022MNRAS.515.6093O} proposed,  for the first time,  two semi-empirical calibrations based on photoionization models to estimate the oxygen abundance of LINERs as a function of intensities of strong optical emission-line ratios. These authors were able to estimate the O/H abundance for 43 LINERs
classified as retired galaxies, i.e., ionized by pAGB stars, finding values in the range $8.5 \lesssim \: 12+\log(\rm O/H) \: \lesssim \: 8.9 $, or  $0.6 \lesssim \: (Z/\rm Z_{\odot}) \: \lesssim \: 1.4$ assuming the solar oxygen value 
$12 + \log(\rm O/H)=8.69$ \citep{Allende_Prieto_2001}.

The determination of the abundance of other elements, such as N and S, is barely found in the literature for LINERs. Indeed, the only available abundance estimation of other heavy elements for LINERS was performed by \citet{2021MNRAS.505.4289P}. In this study was determined the nitrogen-to-oxygen abundance ratio for a sample of 40 LINERs using the \textsc{hii-chi-mistry} code \citep{2014MNRAS.441.2663P, 2019MNRAS.489.2652P}. This code performs a Bayesian-like comparison between the predictions from certain optical emission-line ratios and a large grid of photoionization models that assumes  AGNs as ionization sources.
They found that LINERs present  N/O abundance ratio values similar to SF regions, but lower ($\sim 0.20$ dex) than those of Seyfert~2 nuclei.  The objects considered by \citet{2021MNRAS.505.4289P} were classified as LINERs through two classical emission line ratio diagnostic diagrams,
i.e., [\ion{O}{iii}]$\lambda5007$/H$\beta$ versus [\ion{S}{ii}]($\lambda6716+\lambda6731$)/H$\alpha$ and versus 
[\ion{O}{i}]$\lambda6300$/H$\alpha$.
 However, these diagrams can not 
discriminate the ionization source of the LINERs.
In fact, a gas excited by an AGN,  shocks, or pAGB stars occupies similar regions in classical optical diagnostic diagrams (e.g., \citealt{2021MNRAS.501.1370D, 2022MNRAS.516.5487L, 2023arXiv230102252F}),  requiring additional analysis to carry out galaxy spectral classification (e.g., \citealt{2008MNRAS.391L..29S, 2011A&A...528A..10P, 2011ApJ...736..104J, 2012MNRAS.421.1043S, 2014MNRAS.440.2419R, 2014ApJ...781L..12R, 2017MNRAS.466.2879B, 2017MNRAS.470.4974D, 2018MNRAS.474.1499W, 2019AJ....158....2B, 2019ApJ...876...12A}). It is crucial to note that classical diagnostic diagrams \citep{1981PASP...93....5B,1987ApJS...63..295V} alone can not distinguish the ionization source of LINERs; whether it is due to AGNs, shocks, or pAGB stars. Therefore, an additional analysis is required to determine the ionization source of LINERs.  In this sense, the WHAN diagram proposed by \citet{2010MNRAS.403.1036C}, which takes into account the equivalent width of H$\alpha$ ($\rm EW_{H\alpha}$) versus  [\ion{N}{ii}]$\lambda6584$/H$\alpha$ line ratio, is a useful tool to differentiate the nature of the ionization sources of the objects classified previously as LINER in the BPT diagram, i.e., between evolved low-mass stars (like pAGB stars) and low ionization AGNs.

Nitrogen is predominantly produced in stars with low and intermediate-mass through two distinct nucleosynthetic processes: primary and secondary.  
The primary mechanism was initially proposed by \citet{1971Ap&SS..14..179T}
and later expanded upon by \citet{1974ApJ...190..605T}. In this production mechanism, nitrogen arises from intermediate-mass stars, within the mass range $4 \: \la \: M_{*} \: \la \: 8 \rm \: M_{\odot}$, with a smaller contribution from massive stars ($M_{*} \: \gtrsim \: 8 \rm \: M_{\odot}$) through nuclear reactions involving only the H and He. Additionally, \citet{2002A&A...381L..25M} have suggested that N can also be primarily produced by massive stars with very low metallicity and higher rotation, particularly during the thermal pulse phases of asymptotic giant branch (AGB) stars. On the other hand, secondary nucleosynthesis is explained through the Carbon-Nitrogen-Oxygen (CNO) cycle, where nitrogen is a product of nuclear reactions involving carbon and oxygen. Several studies have explored this secondary process (e.g. \citealt{2023MNRAS.520..782J, 2021MNRAS.508..719G, 2019Sci...363..474J, 2018A&A...610L..16V, 2013MNRAS.436..934W, 2010ApJ...712.1029T, 2003A&A...397..487P, 2000ApJ...541..660H, 1986MNRAS.221..911M}).

One important relation that has been investigated by several authors is the nitrogen-to-oxygen ratio,  as it exhibits a direct correlation 
with variations in metallicity and the star formation history  (e.g., \citealt{2017MNRAS.469.3125C,2019A&ARv..27....3M}). Usually, the \azul{N/O versus O/H metallicity diagram} exhibits a clear behavior: a plateau at low-metallicity regime ($12+ \rm \log(O/H) \: \la \:  8.4$, see e.g., \citealt{ 2013ApJ...765..140A,2019ApJ...874...93B}), that represents the primary nucleosynthesis of the nitrogen, changing abruptly the slope at O/H in the higher metallicity regime, interpreting as the secondary production of the nitrogen (e.g., \citealt{1978MNRAS.185P..77E, 1986MNRAS.221..911M, 2016MNRAS.458.3466V}).  Specifically, as the nitrogen abundance increases with the C and O abundances,  at high metallicities, the nitrogen abundance is expected to evolve quadratically with metallicity, expressed as N/O $\propto$ O/H   or, equivalently, N/H  $\propto$ (O/H)$^{2}$  (see \citealt{2019A&ARv..27....3M}). 

In this study,  the third a series (Paper I - \citealt{2021MNRAS.tmp.1341K} and Paper II - \citealt{2022MNRAS.515.6093O}),  we investigate the nitrogen and oxygen abundances of 43 LINERs. The ionization sources of these LINERs were previously classified as pAGB stars by \citet{2022MNRAS.515.6093O}.   We carried out this analysis using the \textsc{cloudy} code \citep{2017RMxAA..53..385F} to build detailed photoionization models to reproduce strong optical emission line ratios found for the objects in our sample and, thus, estimate the N/H and O/H abundances. We follow a similar methodology as the one applied by \cite{2017MNRAS.468L.113D} to estimate the abundances in a sample of Seyfert~2 nuclei. 
This paper is organized as follows: Section~\ref{dataobs} describes the observational data. In Section~\ref{meth} we present the photoionization model descriptions and the methodology applied to derive the nitrogen and oxygen abundance. In Section~\ref{res} the results and discussions are shown, while in Section~\ref{conc} we present our conclusions.

\section{Observational Data}
\label{dataobs}

To derive the nitrogen and oxygen abundances, observational optical emission-line intensities of 
LINER nuclei taken from the MaNGA survey \citep{2015ApJ...798....7B} were considered. The spectra   comprehend the wavelength range of 3\,600 - 10\,000 \AA, with a spatial resolution of about 2.5 arcsec 
(\citealt{2013AJ....146...32S,2017AJ....154...86W}). 
Auroral emission lines, such as  [\ion{O}{iii}]$\lambda4363$, are not detected. Therefore, our analysis 
exclusively considers only strong emission lines —
[\ion{O}{ii}]$\lambda 3727$, [\ion{O}{iii}]$\lambda 5007$, [\ion{N}{ii}]$\lambda 6584$, [\ion{S}{ii}]$\lambda 6717,6731$, H$\alpha$, and H$\beta$ —  measured with a signal-to-noise ratio higher 3. 
These emission lines have been corrected by reddening using the extinction curve by \citet{cardelli89}, which is widely used as a reference for the internal extinction curve in the local universe galaxies (\citealt{2020ARA&A..58..529S}).
The sample of objects is the same as selected by \citet{2022MNRAS.515.6093O} and is composed of  43 galaxies with LINER emission in their nuclear region and with SF emission in their disks.  A detailed description of the data reduction, reddening correction procedure, etc., are presented in 
\citet{2022MNRAS.515.6093O}. The selection criteria are summarized in what follows.

\begin{figure*}
\includegraphics*[width=1.0\textwidth]{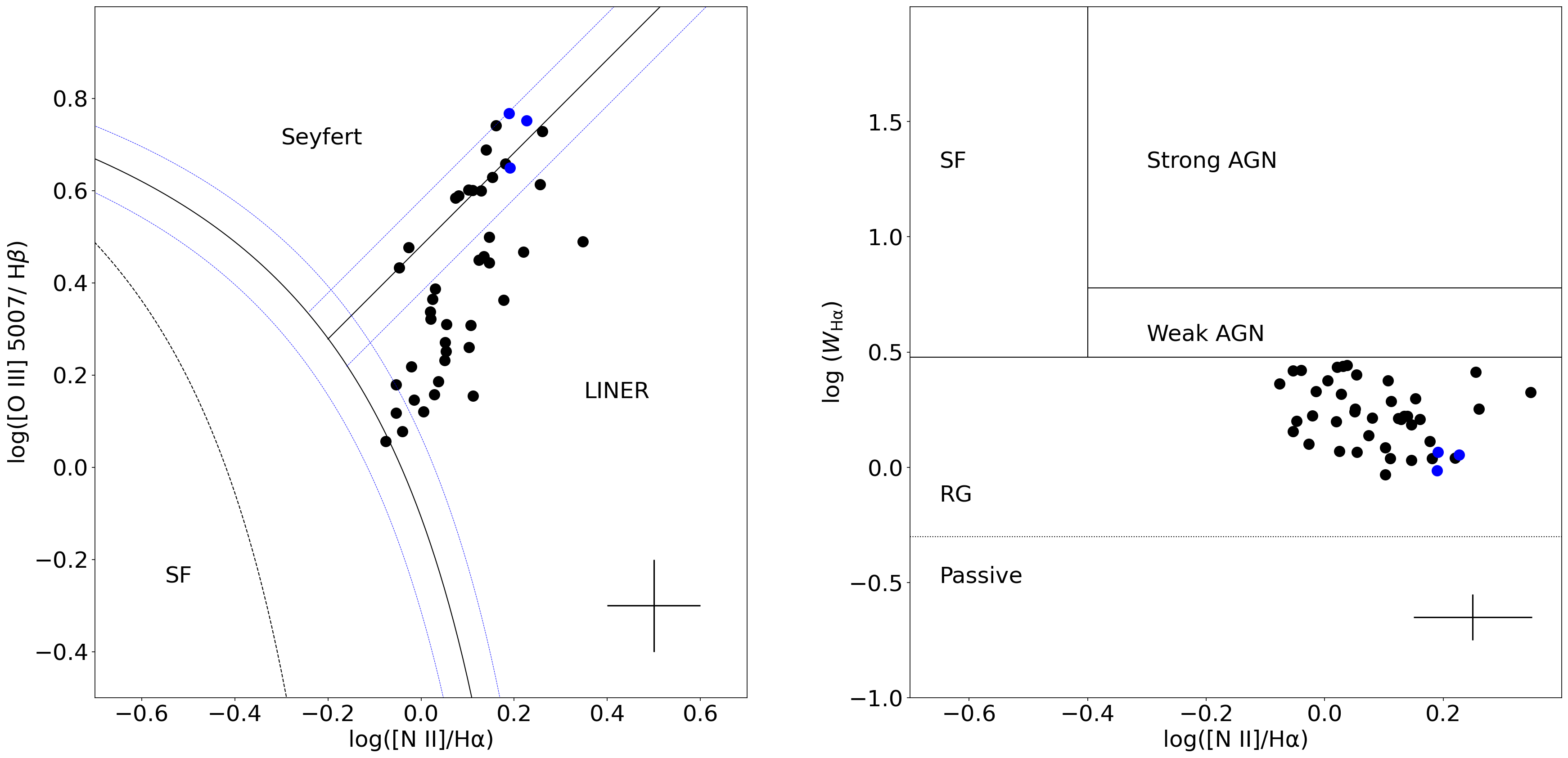}
\caption{Left panel: [\ion{O}{iii}]$\lambda 5007$/H$\beta$ vs. [\ion{N}{ii}]$\lambda 6584$/H$\alpha$ diagnostic diagram. The black solid curve represents the theoretical separation between \ion{H}{ii}-like and AGN-like objects proposed by \citet{kewley01}; the dashed black curve is the empirical star-forming limit proposed by \citet{2003MNRAS.346.1055K}, and the black solid line, proposed by \citet{2010MNRAS.403.1036C}, represents the separation between Seyferts and LINERs. The error bars represent the typical uncertainties of $\pm 0.1$ dex in emission line measurements (see \citealt{2022MNRAS.515.6093O} for a description of fluxes measurements and uncertainties estimations).  Blue dashed lines represent the assumed error of $\pm 0.1$ dex, as explained in Section~\ref{dataobs}. Black points represent the nuclear-integrated flux for each galaxy in our sample. Right panel: WHAN diagnostic diagram. Blue points in both panels are the galaxies  for which we could not reproduce the observed emission line intensities by our models (see Sec.~\ref{res}).}
\label{bpt} 
\end{figure*}

\begin{itemize}
    \item Initially, we used the 
    [\ion{O}{iii}]$\lambda5007$/H$\beta$ versus
    [\ion{N}{ii}]$\lambda6584$/H$\alpha$ diagnostic diagram proposed by
    \citet{1981PASP...93....5B}, known as  [\ion{N}{ii}]-diagram, to classify each spaxel of the individual objects as \ion{H}{ii}-like regions or AGN-like objects. For that, we assumed the theoretical and empirical criteria proposed by \citet{kewley01} and \citet{2003MNRAS.346.1055K}, respectively, and given by: 
    \begin{equation} 
    \label{eq1o}
\rm log([\ion{O}{iii}]\lambda5007/H\beta) \: > \: \frac{0.61}{log([\ion{N}{ii}]\lambda6584/H\alpha)-0.47}+1.19
\end{equation}
 and
    \begin{equation} 
    \label{eq2o}
\rm log([\ion{O}{iii}]\lambda5007/H\beta) \: > \: \frac{0.61}{log([\ion{N}{ii} ]\lambda6584/H\alpha)-0.05}+1.3.
\end{equation}
The AGN-like objects were discerned into Seyfert and LINER categories using the criteria outlined by \citet{2010MNRAS.403.1036C}:

\begin{equation} 
\label{eq3o}
\rm log([\ion{O}{iii}]\lambda5007/H\beta) \: < \:  0.48+1.01\: \times \: log([\ion{N}{ii}]\lambda6584/H\alpha).
\end{equation}

    \item Taking into account the spaxels classified as LINER in the [\ion{N}{ii}]-diagram, we  applied the WHAN [$\rm \log(EW_{H\alpha})$ versus $\log([\ion{N}{ii}]\lambda 6584)$]  diagram (\citealt{2011MNRAS.413.1687C})  to differentiate the nature of the ionization sources of the objects, i.e. between evolved low-mass stars (like post-AGB stars) and low ionization AGNs.  We only selected spaxels classified as Retired Galaxies (RG), i.e.\ $\rm {EW_{H\alpha}} \: < \: 3$ \AA, suggesting that the ionization sources are probably post-AGB stars. 

    \item After the previous classification processes, performed on each individual spaxel, 
    we integrated the fluxes within a circular aperture with a radius of 1\,kpc located at the nuclear region of each galaxy. For these integrated regions we applied the same classification based on the [\ion{N}{ii}]-diagram  and the WHAN  diagrams.

    \item Finally, we selected all the galaxies whose integrated nuclear emission is simultaneously classified as LINER in the [\ion{N}{ii}]-diagram and as RG in the WHAN diagram. In Figure~\ref{bpt} both diagnostic diagrams  are shown. For the [\ion{N}{ii}]-diagram, we considered an error of $\pm 0.1$ dex for the criterion by \cite{kewley01} to separate AGN-like and \ion{H}{ii}-like objects, and the same error value for the criterion by \citet{2010MNRAS.403.1036C} to distinguish Seyfert and LINER objects. 
\end{itemize}

 The stellar mass of the hosting galaxies is in the range of 
$10 \: \la \: \log(M_{*}/\mathrm{M_{\odot}}) \: \la \: 11.2$
and the redshift is in the range of $0.02 \: \la  \: z \: \la \: 0.07$. In Table~\ref{tab1} 
the observational emission-line intensities (in relation to H$\beta$)
of the objects for
which we were able to reproduce their emission-line ratios (see below) are listed.
In the first column of this table is also given the MaNGA identifier "plate-IFU" for each object. 

\begin{table*}\small
\caption{Dereddened fluxes (relative to H$\beta$=1.00)  for our sample of LINER nuclei.
The observed values compiled from the literature are referred as "Obs." while the   predicted values
by the photoionization models as  "Mod." (see Sect.\ref{meth}).}
\label{tab1}
\begin{tabular}{lcccccccccccccc}	 
\noalign{\smallskip} 
\hline 
                                    &    \multicolumn{2}{c}{[O\,II]$\lambda$3727}   &                                    &    \multicolumn{2}{c}{[O\,III]$\lambda$5007 }   &            &      \multicolumn{2}{c}{ [N\,II]$\lambda$6584}     &                        &      \multicolumn{2}{c}{ [S\,II]$\lambda \lambda$6716+31}  &                        &      \multicolumn{2}{c}{ H$\alpha$}\\
\cline{2-3}
\cline{5-6}
\cline{8-9}
\cline{11-12}
\cline{14-15}
Object                          &         Obs.                 &             Mod.                        &                                   &           Obs.                     &           Mod.                 &             &             Obs.                      &              Mod.               &                             &         Obs.                  &  Mod. &                             &         Obs.                  &  Mod.       \\
\hline
7495-12704 & 1.77  &  1.91  && 1.79 &1.70&&  3.25 &3.31 &&2.08 &2.04&&2.87&2.83\\
7977-3704  & 5.06  &  6.08  && 1.65 &1.74&&  2.73 &2.60 &&1.90 &1.67&&2.87&2.85\\
7977-12703 & 3.93  &  4.81  && 1.31 &1.30&&  2.54 &2.40 &&1.88 &1.93&&2.87&2.82\\
7990-12704 & 4.35  &  4.48  && 2.10 &2.09&&  3.01 &3.08 &&3.08 &3.21&&2.87&2.81\\
8083-12704 & 2.27  &  2.54  && 1.32 &1.26&&  2.90 &2.83 &&2.78 &2.56&&2.87&2.82\\
8140-12703 & 8.77  &  9.56  && 1.87 &1.80&&  3.23 &3.37 &&1.89 &1.88&&2.87&2.80\\
8247-3701  & 3.58  &  3.65  && 1.44 &1.52&&  3.06 &3.17 &&1.67 &1.72&&2.87&2.81\\
8249-12704 & 4.25  &  4.07  && 3.88 &4.03&&  3.45 &3.15 &&2.36 &2.38&&2.87&2.81\\
8254-3704  & 5.62  &  5.92  && 2.04 &2.11&&  3.25 &3.41 &&2.22 &2.35&&2.87&2.80\\
8257-1902  & 3.90  &  4.85  && 1.40 &1.38&&  2.77 &2.73 &&1.88 &1.85&&2.87&2.85\\
8259-9102  & 8.45  &  8.29  && 2.44 &2.47&&  3.08 &3.04 &&1.63 &1.62&&2.87&2.79\\
8313-9102  & 5.05  &  4.98  && 2.30 &2.15&&  4.32 &4.49 &&3.04 &3.16&&2.87&2.80\\
8313-12705 & 5.55  &  5.64  && 3.16 &3.17&&  4.02 &4.01 &&3.34 &3.46&&2.87&2.80\\
8318-12703 & 4.19  &  4.85  && 2.77 &2.35&&  4.02 &4.13 &&2.85 &3.03&&2.87&2.81\\
8320-9102  & 4.70  &  4.22  && 2.81 &2.51&&  3.82 &3.75 &&2.30 &2.28&&2.87&2.81\\
8330-9102  & 2.29  &  2.38  && 1.82 &1.77&&  3.63 &3.60 &&2.52 &2.47&&2.87&2.82\\
8332-6103  & 7.97  &  8.23  && 3.84 &4.03&&  3.40 &3.45 &&2.28 &2.39&&2.87&2.79\\
8440-12704 & 3.82  &  4.05  && 1.71 &1.56&&  3.23 &3.24 &&3.25 &2.94&&2.87&2.82\\
8481-1902  & 6.57  &  6.74  && 2.17 &2.23&&  3.00 &3.01 &&2.26 &2.23&&2.87&2.80\\
8482-12703 & 10.04 &  10.40 && 4.10 &4.19&&  5.17 &5.24 &&2.72 &2.71&&2.87&2.79\\
8550-6103  & 4.73  &  5.21  && 1.53 &1.59&&  3.13 &3.22 &&2.59 &2.67&&2.87&2.84\\
8550-12705 & 9.88  &  11.26 && 3.98 &4.13&&  3.71 &3.81 &&1.67 &1.63&&2.87&2.81\\
8552-9101  & 6.49  &  6.15  && 1.43 &1.45&&  3.71 &3.82 &&2.31 &2.43&&2.87&2.81\\
8601-12705 & 3.48  &  4.20  && 1.51 &1.52&&  2.54 &2.54 &&1.55 &1.54&&2.87&2.88\\
8588-9101  & 3.89  &  4.62  && 1.20 &1.19&&  2.62 &2.63 &&1.59 &1.61&&2.87&2.85\\
8138-9101  & 6.30  &  6.18  && 2.32 &2.24&&  3.04 &3.03 &&1.14 &1.14&&2.87&2.81\\
8482-9101  & 6.56  &  6.72  && 5.51 &5.84&&  4.15 &4.09 &&2.16 &2.15&&2.87&2.79\\
8554-1902  & 4.16  &  4.41  && 3.00 &3.08&&  2.70 &2.71 &&2.55 &2.42&&2.87&2.79\\
8603-12703 & 1.86  &  2.00  && 1.14 &1.16&&  2.41 &2.35 &&1.93 &1.83&&2.87&2.83\\
8604-6102  & 5.46  &  5.37  && 2.87 &2.71&&  3.91 &3.82 &&3.59 &3.44&&2.87&2.80\\
8606-3702  & 8.84  &  8.90  && 4.00 &4.29&&  3.63 &3.80 &&3.18 &3.41&&2.87&2.80\\
7990-6103  & 11.40 &  10.90 && 2.71 &2.69&&  2.57 &2.60 &&2.43 &2.50&&2.87&2.85\\
8243-9102  & 10.61 &  10.33 && 3.09 &3.10&&  6.39 &6.51 &&1.97 &2.11&&2.87&2.82\\
8243-12701 & 12.46 &  12.20 && 4.26 &4.31&&  4.08 &4.11 &&1.92 &1.96&&2.87&2.79\\
8332-12705 & 18.37 &  14.63 && 4.46 &4.40&&  4.46 &4.48 &&2.08 &1.97&&2.87&2.83\\
8549-3703  & 16.61 &  14.03 && 4.88 &5.02&&  3.96 &3.77 &&2.77 &3.06&&2.87&2.83\\
8550-12704 & 10.91 &  10.56 && 2.93 &2.95&&  4.77 &4.72 &&3.08 &3.11&&2.87&2.82\\
8138-3702  & 13.32 &  13.66 && 3.98 &4.05&&  3.86 &3.71 &&3.76 &3.56&&2.87&2.80\\
8482-3704  & 10.00 &  9.17  && 2.03 &2.11&&  3.67 &3.61 &&1.95 &2.02&&2.87&2.80\\
8604-12703 & 12.39 &  12.17 && 5.35 &5.29&&  5.23 &5.33 &&2.64 &2.65&&2.87&2.82\\
\hline 
\end{tabular}
\end{table*}

\section{Photoionization models}
\label{meth}
\subsection{Initial parameters}\label{parameters}



Individual photoionization models,  built with the \textsc{cloudy} code version 17.02 \citep{2017RMxAA..53..385F}, were used to reproduce the observed emission line intensities (in relation to H$\beta$) for each object of our sample. Initial parameters for each model were derived following the methodology described below.

\begin{enumerate}

    \item Ionization parameter: the logarithm of the
    ionization parameter ($\log U$) was derived by the equation proposed by \citet{2022MNRAS.515.6093O}:
      \begin{equation*}
        \log U = 0.57 (\pm 0.01) x - 3.19 (\pm 0.01),
    \end{equation*}   
     being $x = \log ([\ion{O}{iii}] \lambda 5007/[\ion{O}{ii}]\lambda 3727)$ obtained from the observational data. A 
     plane-parallel geometry was assumed in the models  (see the discussion by \citealt{2022MNRAS.515.6093O} about the selection of this parameter).

    \item Spectral energy distribution (SED):  we assumed the SED represented by a post-AGB star atmosphere model by \citet{2003A&A...403..709R},  with the logarithm of the surface gravity $\log (\rm g) = 6$ and an effective temperature of $T_{\rm eff}=190$ kK.  

    \item Electron density ($N_{\rm e}$): we assumed a constant  $N_{\rm e}$ value along the nebular radius. For each object,  $N_{\rm e}$ was derived through the observational [\ion{S}{II}]$\lambda6716/\lambda6731$ line ratio, assuming an electron temperature equal to 10\,000 K,  which is a representative value for  \ion{H}{ii} regions (e.g. \citealt{2003ApJ...591..801K, 2020ApJ...893...96B}), and by using the {\sc temden} task of the {\sc iraf} software. It was possible to derive $N_{\rm e}$ for 24 objects. For the remaining of the galaxies in the sample, we were not able to directly derive $N_{\rm e}$ due to the [\ion{S}{II}] line ratio being higher than $\sim 1.4$ (see \citealt{1989agna.book.....O}). Therefore, for the latter, we considered  $N_{\rm e} = 400$ cm$^{-3}$ as the initial value. This value is the average $N_{\rm e}$ obtained from the objects for which we were able to derive $N_{\rm e}$ through the [\ion{S}{II}] line ratio.

    \item Metallicity ($Z$): the  value of the oxygen abundance for each object was derived  through the semi-empirical calibration proposed by \citet{2022MNRAS.515.6093O}:   
    \begin{equation*}\label{eq_n2}
    12 + \log ({\rm O/H}) = 0.71 (\pm 0.03) N2 + 8.58 (\pm 0.01),
    \end{equation*}
     being $ N2= \log ([\ion{N}{ii}] \lambda 6584/\rm H\alpha)$ obtained from the observational data. To convert the oxygen abundance into metallicity we applied the following:
       \begin{equation*}\label{metal}
    (Z/\rm Z_{\odot})=10^{8.69-[12+ \log(O/H)]}
    \end{equation*}
    being 8.69 dex the solar oxygen abundance (\citealt{Allende_Prieto_2001}). Following \citet{2017MNRAS.468L.113D}, models are assumed to be dust-free. The effects of the presence of dust in the gas phase on the inferred N and O abundances are discussed in Sec.~\ref{implications}.     
   It is important to note that the solar abundances of O, N, and S correspond to the default values defined in the {\sc cloudy} version utilized for constructing the photoionization models (Table 7.1 of {\sc cloudy} manual {\sc hazy 1}\footnote{\url{http://web.physics.ucsb.edu/~phys233/w2014/hazy1_c13.pdf}}).
\end{enumerate}

\subsection{Fitting model methodology}\label{fitting}


For each galaxy in our sample, we built an initial model with the input parameters as previously mentioned to reproduce the observational emission line intensities. Then, new models were built  varying, separately, $Z$, N/H, S/H, $T_{\rm eff}$, and $\log U$, considering a step of $\pm 0.2$ dex, the typical uncertainties in nebular parameter estimations derived through photoionization models (see e.g., \citealt{2011MNRAS.415.3616D}).

In particular, some uncertainty in the measured flux of [\ion{S}{ii}] emission lines in our sample
is obtained due to the contribution of the diffuse ionized gas
(DIG, see \citealt{2019MNRAS.485..367K} and references therein), which increases with the decrease of
the H$\alpha$ surface brightness of galaxies (e.g., \citealt{2007ApJ...661..801O, 2017MNRAS.466.3217Z}).
In fact, \citet{2023A&A...669A..88P}, who used the \textsc{HCm-Teff} code \footnote{\url{http://www.
iaa.csic.es/~epm/HII-CHI-mistry.html}} to constraint physical parameters of SF regions, found that diagrams assuming the [\ion{N}{ii}]$\lambda6583$ line (less affected by DIG contribution, e.g., \citealt{2009ApJ...704..842B})  instead of 
[\ion{S}{ii}], leads to infer lower $T_{\rm eff}$ value for the considered objects.   

We assumed that a model successfully represents an object if it reproduces the intensities of emission lines within an uncertainty lower than $\pm25$ per cent.
If no model was able to satisfy this criterion considering the $N_{\rm e}$ value defined as described above,
a new series of models varying $N_{\rm e}$ were built. For 13 (over the 24) objects with $N_{\rm e}$ derived from their observed spectra (using {\sc iraf}) we have to run models varying $N_{\rm e}$ to fit the observed intensities ratios.   These objects are highlighted with the superscript "$(a)$" in Table~\ref{saidas}. This methodology is  similar to that used by \citet{2017MNRAS.468L.113D} to derive the nitrogen and oxygen abundances for a  sample of Seyfert\,2 objects.  

The {\sc phymir} optimization method (\citealt{hoof}), implemented into the
\textsc{cloudy} code \citep{2017RMxAA..53..385F}, was applied to vary the nebular parameters
and select the best model. In Table~\ref{tab1} the model-predicted emission-line intensities ratios for each object of the sample are listed, while the 
resulting 
final model parameters are listed in Table~\ref{saidas}. The  atmosphere models of \citet{2003A&A...403..709R} are defined for
 $T_{\rm eff}$ values between  50 and 190 kK with a step of 10 kK. Thus, the
 $T_{\rm eff}$ values derived for our sample are the ones interpolated between the \citet{2003A&A...403..709R} available values. 

To obtain the uncertainty in our abundance estimations, we consider as benchmarks three models with metallicities $(Z/\rm Z_{\odot})=2.2, 1.0$ and 0.2, which represent the range and the mean value derived for the objects in our sample
(see Sect.~\ref{res}). The 
$T_{\rm eff}$, $N_{\rm e}$ and $\log U$ parameters were assumed to be 115\,000 K, 400 cm$^{-3}$ and $-3.5$, respectively, about the average values obtained from our model fitting to the observational data (see Table~\ref{saidas}). Taking into account these model parameters, 
a series of models varying separately 
the O/H and N/H abundances  by a factor of $\pm0.3$ dex (step of 0.1 dex) were built. Thus, the uncertainty in our fitting was obtained considering the abundance range, adopted in the models,  where the emission line intensities differ by $\pm25$ percent from the originally anticipated values (see also \citealt{2017MNRAS.468L.113D}).
We found that  variations of about $\pm 0.2$ dex and $\pm 0.1$ dex in O/H and N/H abundances, respectively, produce
variations of about $\pm25$ per cent in the intensities of the [\ion{O}{ii}]$\lambda3727$/H$\beta$, [\ion{O}{iii}]$\lambda5007$/H$\beta$,
[\ion{N}{ii}]$\lambda6584$/H$\beta$, and
[\ion{S}{ii}]$(\lambda6716+\lambda6731)$/H$\beta$ emission
line ratios considered.

Since the \textsc{phymir} method implemented in the \textsc{cloudy} \citep{2017RMxAA..53..385F} code does not produce error estimates in the nebular parameters, these values are considered uncertainties
in our O/H ($\pm0.2$ dex) and N/H ($\pm0.1$ dex) abundance estimations.

\begin{table}
\centering
\caption{Assumed model parameter values used to fit the emission lines
observed of our LINER sample.}
\label{saidas}
\begin{tabular}{@{}lccccc@{}}
\hline
Plate-IFU  &  log(O/H) &  log(N/H) & $ T_{\rm eff}$ (K)  & $\log U$ &    $N_{\rm e}$ (cm$^{-3}$)  \\     	     
\hline
7495-12704 & $-3.01$    & $ -3.43$    &85186  & $-3.20$    &121 \\ 
7977-3704  & $-3.67$    & $ -4.36  $    &119218 & $-3.63$    &1615\\ 
7977-12703 & $-3.68$    & $ -4.36 $    &85852  & $-3.63$    &2793$^{(a)}$\\ 
7990-12704 & $-3.09$    & $ -3.82  $    &85187  & $-3.45$    &843$^{(a)}$ \\ 
8083-12704 & $-3.10$    & $ -3.66  $    &80954  & $-3.36$    &41$^{(a)}$	\\ 
8140-12703 & $-2.97$    & $ -3.98  $    &147970 & $-3.76$    &1261$^{(a)}$\\ 
8247-3701  & $-3.05$    & $ -3.73  $    &71565  & $-3.48$    &1025\\ 
8249-12704 & $-3.04$    & $ -3.69  $    &92854  & $-3.20$    &454$^{(a)}$ \\ 
8254-3704  & $-3.09$    & $ -3.87  $    &95126  & $-3.55$    &1100$^{(a)}$\\ 
8257-1902  & $-3.83$    & $ -4.33   $    &92931  & $-3.56$    &675$^{(a)}$ \\ 
8259-9102  & $-3.28$    & $ -4.09  $    &101921 & $-3.55$    &272$^{(a)}$ \\ 
8313-9102  & $-3.19$    & $ -3.74  $    &85510  & $-3.46$    &834$^{(a)}$ \\ 
8313-12705 & $-3.18 $    & $ -3.78  $    &108132 & $-3.37$    &181 \\ 
8318-12703 & $-3.04$    & $ -3.69   $    &100000 & $-3.48$    &833 \\ 
8320-9102  & $-3.04 $    & $ -3.69  $    &84341  & $-3.40$    &1097\\ 
8330-9102  & $-3.06$    & $ -3.50  $    &87876  & $-3.26$    &83	\\ 
8332-6103  & $-3.19$    & $ -4.02  $    &119486 & $-3.49$    &1384$^{(a)}$\\ 
8440-12704 & $-3.00$    & $ -3.72  $    &104736 & $-3.53$    &331 \\ 
8481-1902  & $-3.11$    & $ -3.98  $    &98296  & $-3.57$    &1077\\ 
8482-12703 & $-3.07 $    & $ -3.84  $    &150321 & $-3.55$    &1191\\ 
8550-6103  & $-3.78 $    & $ -4.25  $    &99555  & $-3.55$    &599$^{(a)}$ \\ 
8550-12705 & $-3.22$    & $ -4.11  $    &150761 & $-3.58$    &1843\\ 
8552-9101  & $-3.07$    & $ -3.81  $    &124483 & $-3.66$    &173 \\ 
8601-12705 & $-3.94 $    & $ -4.37   $    &109605 & $-3.51$    &382 \\ 
8588-9101  & $-3.91$    & $ -4.34  $    &88103  & $-3.55$    &14	\\ 
8138-9101  & $-3.02$    & $ -3.88  $    &103309 & $-3.55$    &561 \\ 
8482-9101  & $-3.28$    & $ -3.86   $    &108303 & $-3.24$    &516 \\ 
8554-1902  & $-3.53$    & $ -4.07  $    &65942  & $-3.25$    &1163\\ 
8603-12703 & $-3.04$    & $ -3.63  $    &71404  & $-3.32$    &134 \\ 
8604-6102  & $-3.14$    & $ -3.80  $    &95684  & $-3.43$    &734 \\ 
8606-3702  & $-3.01$    & $ -3.90  $    &141758 & $-3.52$    &1415\\ 
7990-6103  & $-3.47$	  & $ -4.35   $    &161338 & $-3.60$    &89$^{(a)}$	\\ 
8243-9102  & $-3.33$	  & $ -3.91  $    &155345 & $-3.61$    &1463\\ 
8243-12701 & $-3.13$	  & $ -4.00  $    &154776 & $-3.55$    &593$^{(a)}$\\ 
8332-12705 & $-3.23$	  & $ -4.06  $    &189387 & $-3.52$    &189 \\ 
8549-3703  & $-3.26$	  & $ -4.13  $    &189160 & $-3.48$    &191 \\ 
8550-12704 & $-3.39$	  & $ -4.04  $    &149596 & $-3.59$    &239 \\ 
8138-3702  & $-2.96$	  & $ -4.01   $    &188397 & $-3.59$    &1248\\ 
8482-3704  & $-3.35$	  & $ -4.09  $    &121998 & $-3.63$    &103 \\ 
8604-12703 & $-3.30$	  & $ -3.97  $    &167932 & $-3.46$    &363 \\ 
 
\hline
\end{tabular}
\begin{minipage}[c]{1\columnwidth}
$(a)$ Objects for which we needed to build photoionization models assuming as input $N_{\rm e}$ values differ from those derived from observational [\ion{S}{ii}]$\lambda6716/\lambda6731$ line ratio, as described in Sect.~\ref{fitting}.
\end{minipage}
\end{table}

\section{Results and Discussions}
\label{res}

We were able to obtain detailed photoionization models for 40 galaxies from our original sample of 43 objects. The cause of the model's failure in 3/43 LINER models was the difficulty in reproducing both high observational [\ion{O}{ii}]/H$\beta$  and [\ion{O}{iii}]/H$\beta$ values, i.e. values higher than $\sim 15$. These objects
(not listed in Table~\ref{tab1}) are 8131-9102, 8252-12702 and 8258-12704 and they are represented in Fig.~\ref{bpt} by blue points.
Possibly, for these three LINERs, the ionizing source is  hotter  than the upper limit of 190 kK of the \citet{2003A&A...403..709R}
atmosphere models, e.g., accreting, nuclear-burning white dwarfs (WDs) with photospheric temperatures of  $10^{5}$-$10^{6}$ K
\citep{2013MNRAS.432.1640W}.  However, 
\citet{2016MNRAS.461.4505J} argued that the ionizing source in RGs is mainly composed of pAGB stars rather than accreting, nuclear-burning WDs.

\begin{figure*}
\includegraphics*[width=.75\textwidth]{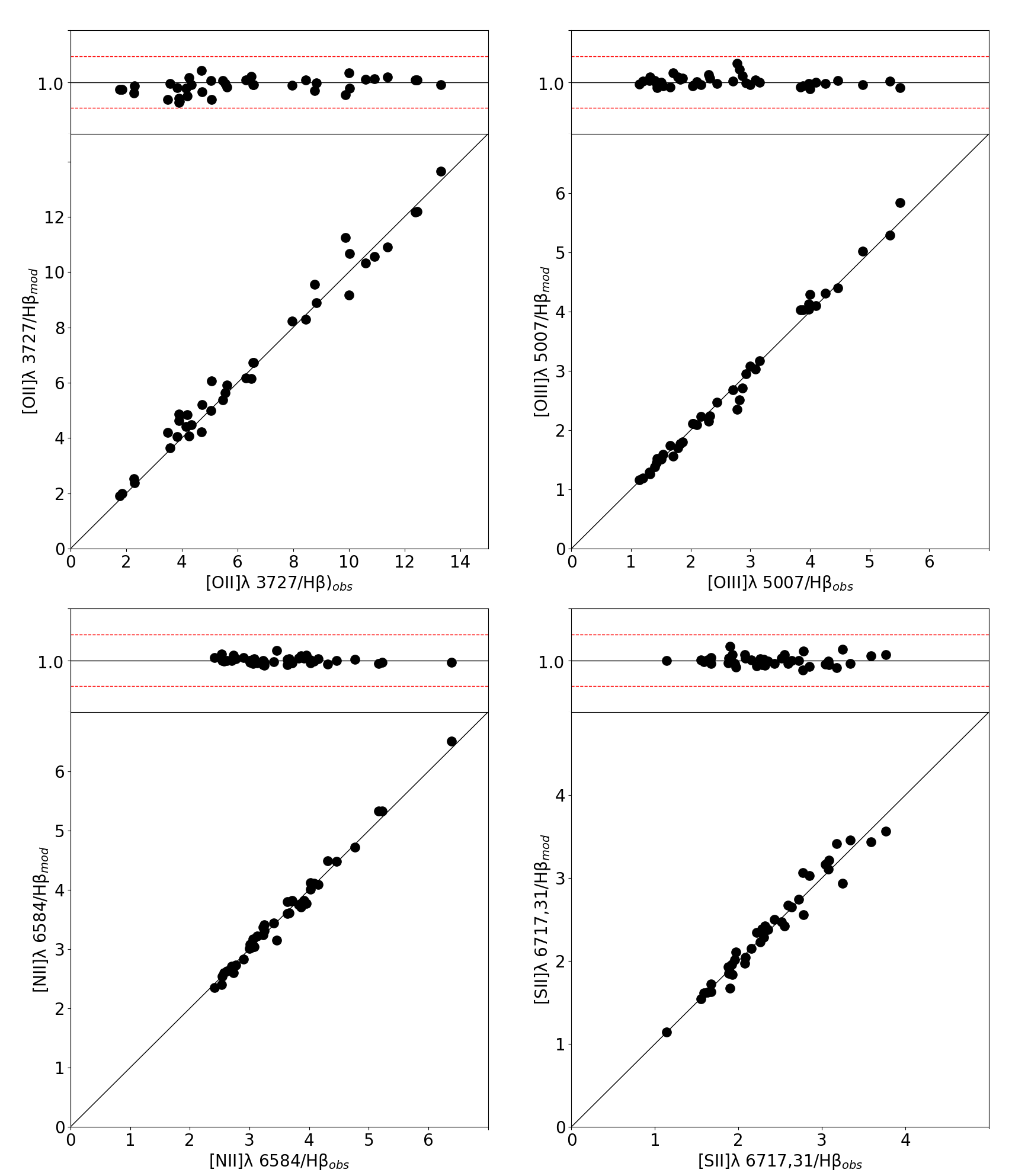}
\caption{Bottom part of each panel: comparison between model predictions ($y$-axis) and observed ($x$-axis) emission line ratios (normalized by H$\beta$) for our LINER sample. 
The solid line represents the equality of the two estimates. 
Top part of each panel: the ratio between the observed and predicted intensities versus the observed emission line intensities ratios. The red dashed lines represent the adopted limit of $\pm25\%$ between the observed and predicted line ratios to consider that a model successfully represents an object.}
\label{ratio} 
\end{figure*}

Fig.~\ref{ratio}, bottom panels, shows the intensities of observational emission-line ratios  ($x$-axis) compared with those predicted by the photoionization models ($y$-axis). In the upper part of each panel of this figure,
the ratios between the observed and predicted intensities versus the observed ones are shown.
It can be noted a very good agreement between the predicted and observational emission line ratios. 

\subsection{Implications of model assumptions}\label{implications}

Before analyzing the results, we discuss some implications
of our model assumptions on the N and O abundance estimates.

\subsubsection{Electron density variation}

 Spatially resolved observational studies have found
that $N_{\rm e}$ varies along the
nebular radius, in the sense that near
the ionizing source higher values are
derived in comparison to those in the outskirt gas regions. For instance, in their pioneering study,
\citet{1959ApJ...129...26O} found, based on the [\ion{O}{ii}]$\lambda 3716/\lambda3729$ line ratio, a steep radial density gradient in the Orion Nebula, with $N_{\rm e}$  decreasing from $\sim 10^{4}$\,cm$^{-3}$
in the centre to $\sim10^{2}$\,cm$^{-3}$ near the edge. Additional studies have confirmed the existence of radial profiles or variations of $N_{\rm e}$ in SF regions (e.g., \citealt{1992A&A...260..370C, 2000A&A...357..621C, 2010MNRAS.405.2651M, 2011MNRAS.417..420M, 2013ApJ...765L..24C, 2017MNRAS.464.4835O}) and AGNs
(e.g., \citealt{ 2017MNRAS.471..562C, 2018MNRAS.476.2760F, 2018A&A...618A...6K, 2019A&A...622A.146M}). Moreover,
$N_{\rm e}$ estimation based on lines emitted by ions with higher ionization potential than $\rm S^{+}$, such as  [\ion{Ar}{iv}] and [\ion{Fe}{iii}] emission-lines, which trace the density in the inner gas layers,  have revealed that $N_{\rm e}$ in SF regions (e.g., \citealt{2011MNRAS.417..420M}) and AGNs (e.g., \citealt{2017MNRAS.471..562C}) can reach values in order of 20\,000 cm$^{-3}$. 
Such high $N_{\rm e}$ values can produce a suppression of some forbidden emissions lines and a wrong interpretation from photoionization models, hence the density assumed in our models is derived from the observational [\ion{S}{ii}] lines ratio, which could not be a representative value for the entire gas. Moreover, our analysis considers   
integrated  spectra of the nuclear zones of the galaxies, therefore, the electron density
values obtained in the present work must be considered as mean
values.   

To investigate the  $N_{\rm e}$ influence on our abundance estimations, we carried out some simulations. Firstly, we considered
the representative model for
the 8588-9101 object as a benchmark, hence it presents the lowest derived $N_{\rm e}$ value (14 cm$^{-3}$; see Table~\ref{saidas}). Thus, we carried out a new fitting to the observational line intensities ratios of this object (see Table~\ref{tab1}) assuming six fixed values of $N_{\rm e}$: 1\,000, 2\,000, 3\,000, 4\,000,  5\,000 and 10\,000 cm$^{-3}$. After running these models, we did not observe any abundances variation for models with $N_{\rm e}$= 1\,000, 2\,000, 3\,000 and 4\,000, while for the photoionization model with  $N_{\rm e}= 5\,000$ cm$^{-3}$ we found a discrepancy of approximately $0.15$ dex 
for O/H and around $0.03$ dex for 
N/H when comparing with the benchmark model results.  
We did not find a solution for the model with $N_{\rm e}= 10\:000$ cm$^{-3}$ because it underpredicts the [\ion{O}{ii}]$\lambda3727$/H$\beta$ in comparison to the observational value, possibly due to effects of collisional deexcitation (no present in the 8588-9101 object) in which produce a suppression in the emission of collisionally excited lines. The critical density for the [\ion{O}{ii}]$\lambda 3726$ and $\lambda3729$ lines is in order of $10^{3}$ cm$^{-3}$ (see e.g., \citealt{2023MNRAS.521.1969D}).

Secondly, we assumed a density profile for the gas. For instance, \citet{2018ApJ...856...46R}
found for the AGN in Mrk\,573 a peak of about
$N_{\rm e}=$3\,000 cm$^{-3}$  at the center and a decrease following a shallow power
law  with the radial distance ($r$):
$N_{\rm e} \: \sim \: r^{\alpha}$, being $\alpha=-0.5$. The power law derived by these authors includes only points inner $  \sim1.7$ arcsec ($\sim 600$ pc) of radius (see also \citealt{2018ApJ...867...88R, 2021ApJ...910..139R, 2021MNRAS.507...74R}). 
Regarding LINERs, it seems that such density profile found for SF regions and AGNs is also
present. In fact, \citet{2015ApJ...814..149C} presented an analysis of optical emission-line intensities, obtained by
ground-based and  Hubble Space Telescope spectroscopy, of a sample of $\sim100$ nearby galaxy nuclei, including LINERs. These authors found, for all types of nuclei, indications that gas densities are generally
higher in the most central regions.
To analyze the effect of density profile on our abundance estimates, we
choose as a benchmark the model representing the 7977-12703 galaxy, hence the highest density value ($N_{\rm e}= 2793$ cm$^{-3}$, see Table~\ref{saidas}) was derived for this object. We assumed $\alpha=-2.0, -1.0, -0.5$, and $-0.1$ and carried out new fitting to the observational line ratio intensities of this object. We found that  
models assuming radial density profiles result in very similar line intensity ratio (less than 10\%) as those predicted by the constant density model and, consequently, the same nebular parameters as those for the constant density model ($N_{\rm e}= 2793$ cm$^{-3}$). The same result was found by \citet{2018MNRAS.479.2294D},
who used photoionization model simulations of narrow-line regions of AGNs. 

Therefore, we can assume that the uncertainties due to high and constant electron density values
for our O/H and N/H estimates are  $\sim0. 15$ and $\sim0. 03$ dex, respectively. Also, we conclude that density variation along the nebular radius does not influence our abundance estimates (see also \citealt{2020MNRAS.496.3209D, 2023MNRAS.520.1687A}).


\subsubsection{Dust content}

\noindent The abundance of elements (e.g. Fe, Mg, Si, etc.) contained in dust grains in the gas phase of gaseous nebulae is poorly known due to the difficulty of estimating it (see e.g., \citealt{1994ApJ...430..650S, 1995ApJ...449L..77G, 2013MNRAS.432.2112B, 2016A&A...585A.112M, 2023arXiv231007709K}). 
Interstellar dust grains absorb and scatter the ionizing radiation, decreasing the hardness of the radiation field and thus
influencing the temperature and ionization structures of the gas (\citealt{1982A&A...106....1M, 2001MNRAS.323..887C, 2004ApJS..153....9G, 2016MNRAS.456.3354F}).
Also, some fraction of oxygen\footnote{For a review on dust abundance
see \citet{1987ASSL..134..533J, 1989IAUS..135...23J}.} (about
$\sim0.1$ dex) is expected to be locked into dust grains
(see e.g., \citealt{1998MNRAS.295..401E, 1998ApJ...493..222M, 2010ApJ...724..791P}). Thus, any oxygen abundance estimates of the gas phase in gaseous nebulae trend to
underestimate the abundance of this element   if a correct depletion is not taken into account (see e.g., \citealt{2007MNRAS.376..353P}) 

Our models are dust free, thus, we built another model simulation using {\sc cloudy}  code \citep{2017RMxAA..53..385F} taking into account the dust present in the gas phase to analyze the possible effect on our abundance estimates.  Since the dust abundance seems to be correlated with the gas phase
metallicity \citep{2010ApJ...724..791P}, we consider as benchmarks three models
with the highest, intermediate, and lowest metallicity derived from our simulations, i.e. the models 
representing the 8138-3702 [($Z/\rm Z_{\odot})=2.2$], 8549-3703 [($Z/\rm Z_{\odot})=1.0$] and
8601-12705 [($Z/\rm Z_{\odot})=0.2$] objects, respectively (see Table~\ref{saidas}).
Following \citet{2011MNRAS.415.3616D},   the grain abundances \citep{2001ASPC..247..363V} were linearly scaled with the metallicity. 
To take the depletion of
refractory elements onto dust grains into account, the abundance
of the elements Mg, Al, Ca, Fe, Ni, and Na were reduced by a factor
of 10, and Si by a factor of 2 \citep{1995ApJ...449L..77G}, relative to the
adopted metallicity in each model.
The dusty models were fitted to the corresponding observational data of each object and the resulting abundances were compared with those predicted by the dust-free models. From these simulations, we found
that O/H abundances predicted by dusty models are lower by a factor of 0.20, 0.10, and 0.05 dex for the models with the highest, intermediate, and lowest metallicity, respectively, in comparison to predictions by dust-free models. 
The nitrogen abundances did not change taking into account the dust presence in the models.

We define the uncertainty in O/H abundances due to the dust present in our objects as being 0.12 dex, the mean value of the discrepancy derived above. 

\subsection{N and O abundances}
 \label{n-o section}
 The methodology presented by \citet{1969BOTT....5....3P}
 to calculate electron temperatures and elemental abundances of gaseous nebulae opened a new window in astronomy (see also \citealt{1960MNRAS.120..326S, 1968nim..book..483A, 1970QJRAS..11..199O, 1974agn..book.....O}), in the sense that it enabled the knowledge of the chemical abundances in galaxies relying on a (relatively) precise method, i.e. the $T_{\rm e}$-method (see e.g., \citealt{2003A&A...399.1003P,2006MNRAS.372..293H,2008MNRAS.383..209H,2017MNRAS.467.3759T}). Over decades, the use of the $T_{\rm e}$-method has permitted to calculate  heavy elemental abundances (e.g., O, N, S) in nearby galaxies (e.g., \citealt{1975ARA&A..13..113P, 1979ApJ...233..888T, 1982ApJ...255....1R, 1986PASP...98.1032F, 1998AJ....116.2805V, 1999ApJ...511..639I, 2003ApJ...591..801K, 2011MNRAS.414..272H, 2012MNRAS.422.3475H, 2015ApJ...806...16B, 2022ApJ...939...44R}) and, recently, in the early universe
 (see e.g. \citealt{2022ApJ...940L..23A, 2023arXiv230308149S, 2023MNRAS.518..425C}).

Due to the difficulty of using this method, i.e. the need to measure
weak auroral emission lines such as  [\ion{O}{iii}]$\lambda4363$, made it necessary to develop strong-line methods. The method proposed by \citet{1979MNRAS.189...95P} permits to estimate the O/H abundance measuring only the 
$R_{23}$=([\ion{O}{ii}]$\lambda3727$+[\ion{O}{iii}]$\lambda4959+\lambda5007$)/H$\beta$ strong lines ratio.
Further authors have adapted this method to estimate the relative abundances of other elements. For instance, \citet{2005MNRAS.361.1063P} proposed
a calibration between the strong 
$N2O2$=[\ion{N}{ii}]$\lambda6584$/[\ion{O}{ii}]$\lambda3727$ lines ratio and the N/O abundance for SF regions. 
Unfortunately, most studies have predominantly concentrated on developing strong-line methods to estimate ionized gas content in SF regions, while AGNs and LINERs are rarely analyzed due to their poorly understood ionization mechanisms, particularly in the case of LINERs. Thus, there have been few studies estimating gas-phase abundances in regions with ionization similar to AGNs and LINERs, such as \citet{2019MNRAS.485..367K}, \citet{2022MNRAS.513..807D} and \citet{2022MNRAS.514.4465M}.

Recently, \citet{2022MNRAS.515.6093O} proposed the first (semi-empirical) calibration for LINERs between strong emission lines and the O/H abundance.  These authors were able to estimate the abundances for 43 LINERs, possibly ionized by pAGB stars, finding values in the range of $\rm 8.5\: \la \: 12+\log(O/H) \: \la \: 8.8$, or $0.60 \: \la \:  (Z/\rm Z_{\odot}) \: \la  \: 1.40$. \cite{2021MNRAS.505.4289P} compiled optical spectroscopic data of a sample of 40 LINERs from the Palomar Survey \citep{1995ApJS...98..477H,1997ApJS..112..315H} and 25 observed by \citet{2016MNRAS.462.2878P} at the Calar Alto
Observatory, and applied the \textsc{hii-chi-mistry} code (\citealt{2014MNRAS.441.2663P}, hereafter {\sc HCm} code) to derive O/H and N/O. 
This code establishes a Bayesian-like comparison between
the predictions from a grid of photoionization models built with the \textsc{cloudy} code \citep{2017RMxAA..53..385F} covering a large range of input parameters and using the lines emitted by the ionized gas.
These authors found a range of  $\rm 8.0\: \la \: 12+\log(O/H)\: \la \: 8.9$ (with a median  
value of $\sim 8.6$ dex) and  $\rm -0.2\: \la \: \log(N/O)\: \la \: -1.1$ (with a median value of $\sim -0.6$) for their LINERs sample. The nitrogen abundance estimates by  \cite{2021MNRAS.505.4289P} seem to be the first estimation in LINERs. In particular, the N/O abundance ratio is a useful tool to study the interplay of galactic processes, for instance, star formation efficiency, the timescale of infall, and outflow (e.g., \citealt{2023MNRAS.520..782J, 2018A&A...618A.102M}) as well as it can be used as a metallicity indicator (e.g., \citealt{2009MNRAS.398..949P}).

As indicated by the WHAN diagram
(see Fig.~\ref{bpt}), the LINERs sample selected for the
present work has pAGB stars as the main ionization source, allowing to apply of the $T_{\rm e}$-method and/or standard photoionization models to obtain elemental abundances. However, for our LINERs sample, auroral lines were not measured and it was only possible to estimate both O and N abundances through detailed photoionization models. This technique has also been employed in SF regions  (e.g., \citealt{2006A&A...452..473D, 2010MNRAS.404.2037P}) and AGNs (e.g., \citealt{2017MNRAS.468L.113D, 2021MNRAS.501.1370D, 2017MNRAS.469.3125C}) studies and trends
to overestimate elemental abundances by  0.1-0.2 dex in comparison to those through the $T_{\rm e}$-method (e.g., \citealt{2014MNRAS.441.2663P, 2020MNRAS.496.3209D}). The source of this discrepancy is an open problem in astronomy and it is been attributed to 
the presence of electron temperature fluctuations in gaseous nebulae \citep{1967ApJ...150..825P}
and AGNs \citep{2021MNRAS.506L..11R}, 
departure from Maxwell–Boltzmann equilibrium
energy distribution \citep{2012A&A...547A..29B, 2012ApJ...752..148N} or a simplified  geometry (e.g., \citealt{2022ApJ...934L...8J}) assumed in photoionization models. However, the scenarios listed above do not provide a plausible explanation for this discrepancy problem.
It is beyond the goal of this paper to address this issue and we only emphasize that our theoretical abundances can be possibly overestimated by 0.1-0.2 dex in comparison to those via the $T_{\rm e}$-method.

\begin{figure}
\includegraphics*[width=0.49\textwidth]{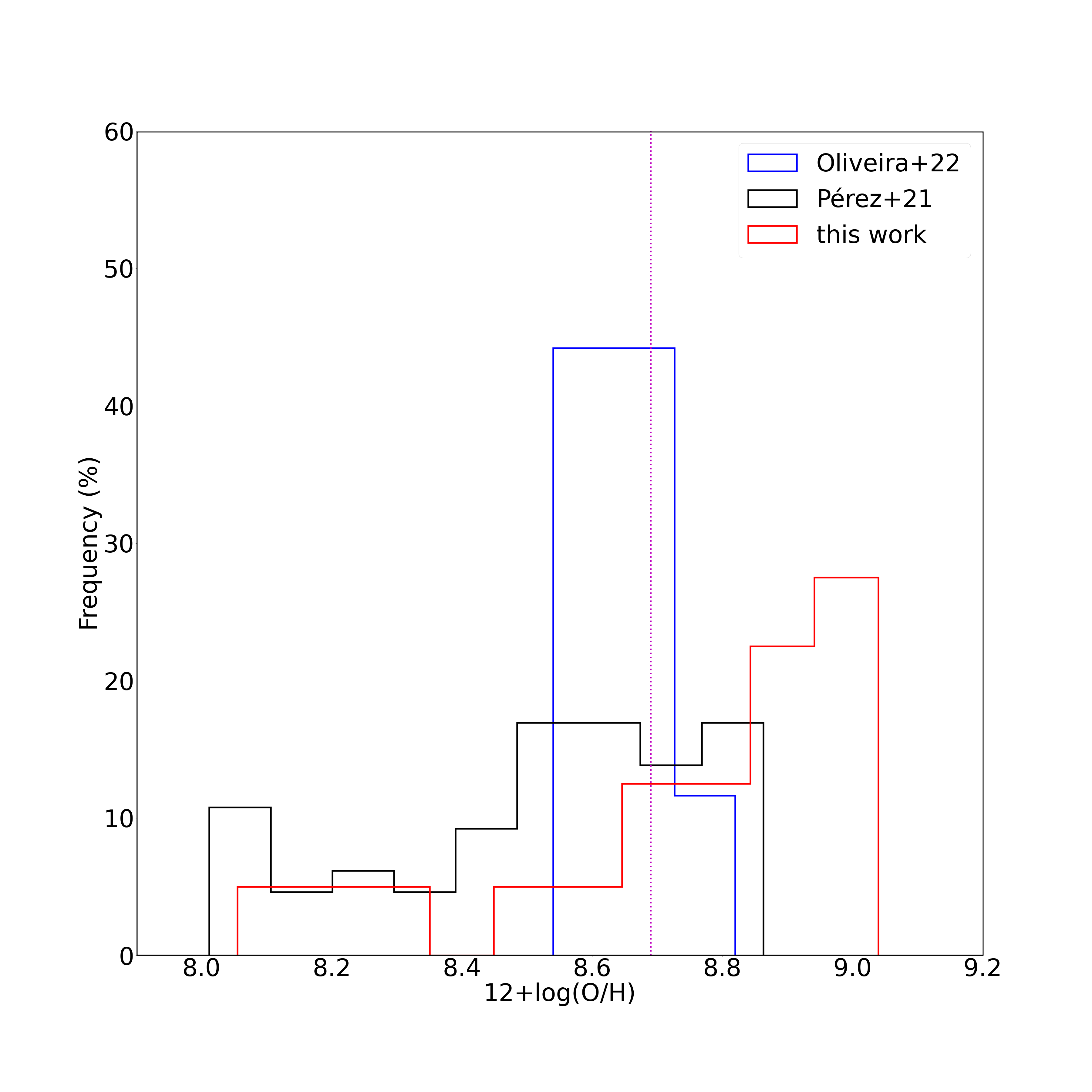}
\caption{Histograms comparing  O/H estimations derived by \citet{2022MNRAS.515.6093O}, as well as the estimations derived by \citet{2021MNRAS.505.4289P} and O/H values derived through our detailed models. The dashed vertical line represents the solar abundance  $\rm 12+\log(O/H)_{\odot} =8.69$, taken from \citet{Allende_Prieto_2001}.}
\label{histogram_oxy} 
\end{figure}

Regarding the range of the derived oxygen abundances, we found a distribution
ranging from $\rm 8.0\: \la \: 12+\log(O/H) \: \la \: 9.0$
$[0.20 \: \la \: (Z/\rm Z_{\odot}) \: \la \: 2.0]$, with a mean value of <12+log(O/H)>=$8.83\pm0.27$. 
In Fig.~\ref{histogram_oxy}, we show a histogram of the computed oxygen abundance distribution.
Notably, $\sim72$\% of the objects in our study present values higher than the solar abundance.
The resulting O/H range is  wider than that  found by \citet{2022MNRAS.515.6093O}, --  $\rm 8.5\: \la \: 12+\log(O/H) \: \la \: 8.8$ or
$0.6 \: \la \: (Z/\rm Z_{\odot}) \: \la \: 1.4$ (blue histogram in Fig. \ref{histogram_oxy}),
obtained from a semi-empirical calibration, and assuming a fixed of (N/O)-(O/H) abundance relation taken from \citet{2020MNRAS.492.5675C}.
On the other hand, the found range is consistent with that derived by \cite{2021MNRAS.505.4289P} {(black histogram in Fig. \ref{histogram_oxy})},  relying on the {\sc HCm} code, taking into account the uncertainty in the O/H estimations via the present method ($\sim 0.2$ dex).


\begin{figure}
\includegraphics*[width=0.49\textwidth]{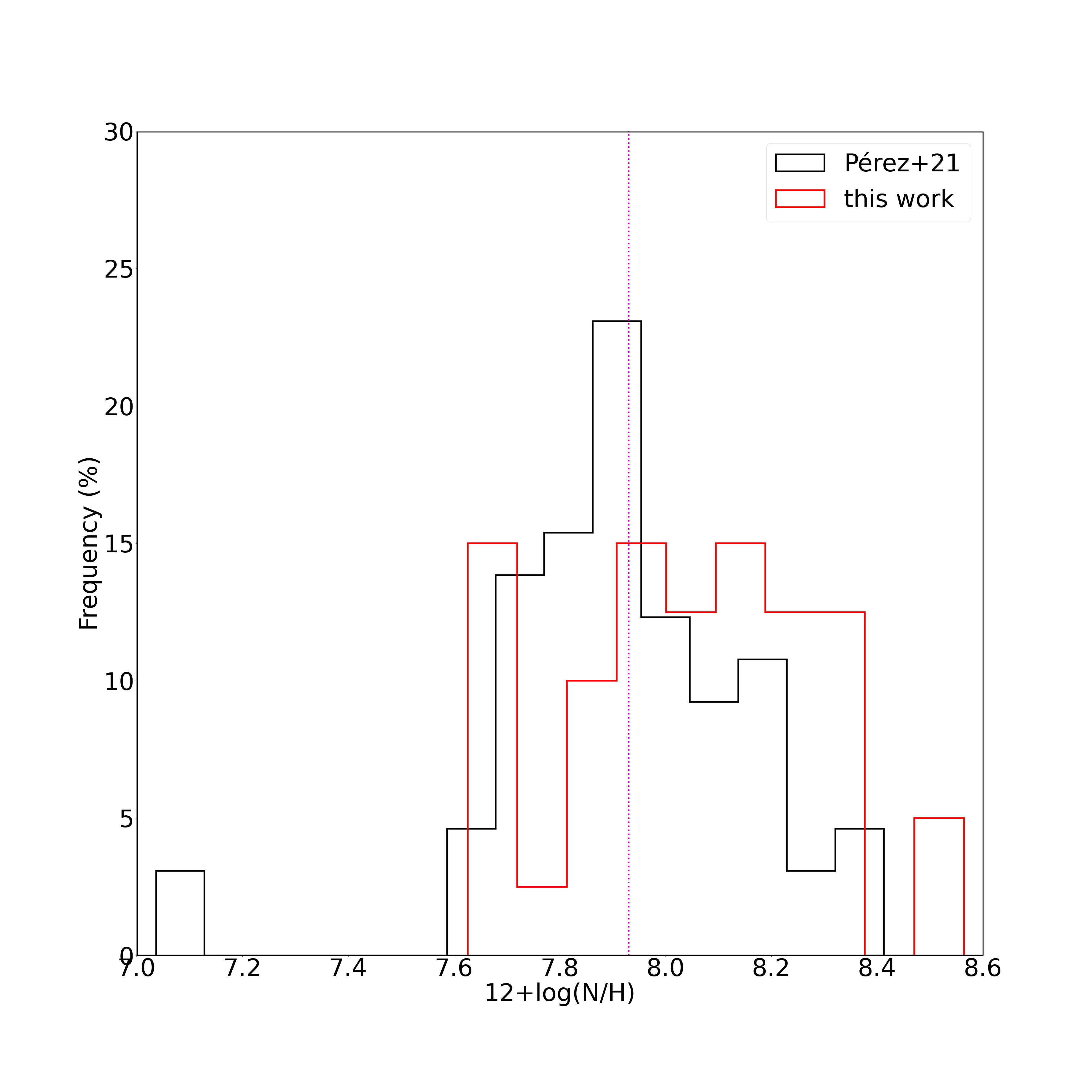}
\caption{Histograms comparing  N/H estimations derived by \citet{2021MNRAS.505.4289P} and N/H values derived through our detailed models. The dashed vertical line represents the solar abundance  $\rm 12+\log(N/H)_{\odot} =7.93$, taken from \citet{2001AIPC..598...23H}.} 
\label{histogram_nitrogen} 
\end{figure}

We found that our LINERs sample shows a range of  $\rm 7.6 \: \la \: 12+\log(N/H) \: \la \: 8.5$ or
$\rm 0.4 \: \la \: (N/N_{\odot}) \: \la \: 3.7$. The mean value is $\rm <12+\log(N/H)>=8.05\pm0.25$ or $\rm <(N/N_{\odot})>\sim1.6$, having oversolar values in about  70 per cent of the objects. The nitrogen abundance range derived via the {\sc Hcm} code by \cite{2021MNRAS.505.4289P}
is $\rm 7.04 \: \la \: 12+\log(N/H) \: \la \: 8.4$ [$\rm 0.1 \: \la \: (N/N_{\odot}) \: \la \: 3.0$], in agreement with our results, as shown in Fig.~\ref{histogram_nitrogen}. However, the mean value derived by 
\cite{2021MNRAS.505.4289P} is
$\rm <12+\log(N/H)>=7.93\pm0.24$ [$<\rm (N/N_{\odot})>\sim1.0$], a factor of $\sim0.1$ dex lower than the mean value obtained for our sample.
The reason for this discrepancy is due to three outlier objects of the \cite{2021MNRAS.505.4289P} 
sample present very low ($\rm 12+\log(N/H) \: \la \: 7.10$) nitrogen abundances. 

\begin{figure}
\includegraphics*[width=0.49\textwidth]{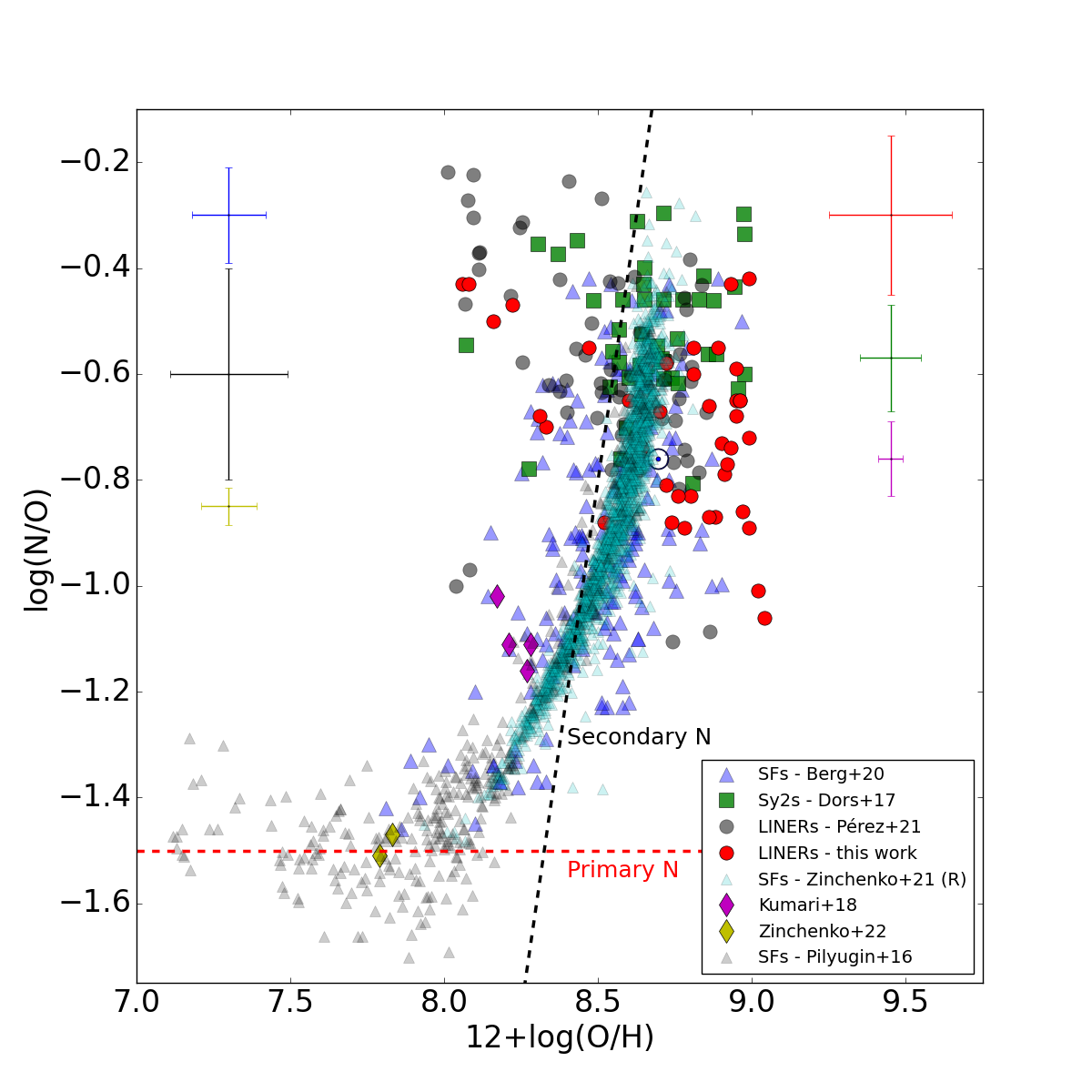}
    \caption{log(N/O) vs. 12+log(O/H). Red points are the predicted values for our LINERs sample. Blue triangles are the values estimated by \citet{2020ApJ...893...96B} for disk \ion{H}{ii} regions. 
    Green squares are the values derived by \citet{2017MNRAS.468L.113D} for Seyfert~2 nuclei, while grey points are the abundances reported by \citet{2021MNRAS.505.4289P} for 65 LINERs. Cyan triangles are values derived applying the R calibration \citep{2016MNRAS.457.3678P} to 1431 SF galaxies as reported by \citet{2021A&A...655A..58Z}. 
    Purple diamonds are the \citet{2018MNRAS.476.3793K} data, and yellow diamonds are the \citet{2022AN....34320048Z} ones. Grey triangles are the SFs sample by \citet{2016MNRAS.457.3678P}.
    The solar values taken from \citet{2001AIPC..598...23H} and \citet{1998SSRv...85..161G} are also included in the plot with the solar symbol. The red and black dashed lines are taken from \citet{2015MNRAS.449..867B} and represent the expected N/O when nitrogen is mainly due to a primary production and a linear fit to the N/O distribution of galaxies in SDSS DR7 with $\rm 12+log(O/H)>8.3$ (secondary nitrogen production), respectively. Error bars represent the average uncertainties in O/H and N/O estimations.}
\label{comparation} 
\end{figure}

 To explore a possible correlation between nitrogen and oxygen abundances, we plot in Fig.~\ref{comparation} the $\rm \log (N/O)$ versus $\rm 12+\log(O/H)$ estimations for the objects in our sample. In this figure, we also  included:

\begin{enumerate}
\item Estimates for 65 LINERs  compiled from the dataset presented in \cite{2021MNRAS.505.4289P}.   These authors determined the abundances using the \textsc{HCm} code for a sample of 40 LINERs in the local universe and taken from the Palomar Survey \citep{1995ApJS...98..477H,1997ApJS..112..315H}. Additionally, they compared their results with a sample of 25 more distant LINERs at 
$z = 0.04 - 0.11$ observed by \citet{2016MNRAS.462.2878P} at the Calar Alto Observatory.

\item Abundance values of 176 disk \ion{H}{ii} regions located in four local spiral galaxies derived using the $T_{\rm e}$-method by \citet{2020ApJ...893...96B} and resulting from the CHAOS project \footnote{\url{https://www.danielleaberg.com/chaos} \citep{2015ApJ...806...16B}}.
    
\item Estimates for 44 Seyfert~2 nuclei ($z <0.1)$ derived by \citet{2017MNRAS.468L.113D}  through detailed photoionization modeling.
\item Nuclear abundance values of 1431 spiral galaxies obtained extrapolating their disk metallicity gradients derived by \citet{2021A&A...655A..58Z}, by applying the R calibration \citep{2016MNRAS.457.3678P}, 
 observed by the MaNGA survey.


\item Direct T$_e$ estimates for 4 \ion{H}{ii} belonging to the dwarf blue compact galaxy NGC\,4640 derived by \cite{2018MNRAS.476.3793K}.

\item  Estimates for 2 \ion{H}{ii} regions of the dwarf irregular galaxy galaxy NGC\,4163 by \cite{2022AN....34320048Z}.

\item Estimations for \ion{H}{ii} regions derived by \citet{2016MNRAS.457.3678P} using the C method \citep{2012MNRAS.424.2316P}. These objects reach lower oxygen abundance values showing the change between the primary and secondary nitrogen production behaviour \citep[see also][]{2017MNRAS.468L.113D}.

\item 
The primary  (red dashed line) and secondary production (black dashed line) lines, as proposed by \citet{2015MNRAS.449..867B}. In regions with low metallicities, i.e. $\rm 12+\log(O/H) < 8.35$, nitrogen is considered as a primary nucleosynthetic product, showcasing a flat behaviour with $\rm \log(N/O) \sim -1.5$. Conversely, for regions with $\rm 12+\log(O/H) > 8.35$, nitrogen takes on the role of a secondary nucleosynthetic product. In this context, a 
relationship between N/O and O/H is expressed as $\rm \log (N/O) = 4.0 \times [\rm 12+\log(O/H)] -8.7$, as derived by \citet{2015MNRAS.449..867B} using data obtained 
through the Sloan DR7.
\end{enumerate}

In Fig.~\ref{comparation}, we can note that LINERs present similar N/O and O/H abundances to those derived for Seyfert 2 nuclei,   indicating that these distinct
object classes have similar ISM enrichment. Moreover, LINERs estimates occupy a region similar to the most metallic and innermost disk \ion{H}{ii} regions. This is an expected result because most parts of spiral galaxies exhibit negative radial abundance gradients, with the highest abundances derived at the lowest galactocentric distances (see e.g., \citealt{1998AJ....116.2805V, 2003ApJ...591..801K, 2004A&A...425..849P, 2005A&A...437..837D, 2022ApJ...939...44R, 2022MNRAS.513..807D}). In Table~\ref{liners},  the statistics of the estimates shown in Fig.~\ref{comparation} are presented. As can also be seen in Fig.~\ref{comparation},  our LINERs estimates are located toward the higher N/O regions, having an orthogonal distribution (with a correlation coefficient of $r=-0.45$) with respect to the values predicted by the line representing the secondary production of nitrogen. In our sample, LINERs with the highest O/H values present the lowest N/O ratios, which are, in turn, lower than those expected from the theoretical secondary curve and \ion{H}{ii} region observations. It could be indicating that our LINERs present mainly a secondary
nitrogen production together with 
different mechanisms producing the observed offset, as we will discussed in what follows.

We can also note in Fig.~\ref{comparation} that a distinct (N/O)-(O/H) relation for LINERs in comparison to those for other object classes is observed. To confirm that, we consider the abundance estimates in 
Fig.~\ref{comparation} and conducted a linear fitting analysis (not shown) to the abundance estimates of the samples, except for the sample of SF galaxies by  \citet{2021A&A...655A..58Z}. For this specific SF galaxy sample,  we employed a quadratic equation for the fitting process.  The fitted equations and the  Pearson correlation coefficient ($r$) values are listed in  Table~\ref{liners}. As can be seen, a positive correlation 
between log(N/O) and  12+log(O/H) is clearly found for the SF galaxy samples from \citet{2021A&A...655A..58Z} and \citet{2020ApJ...893...96B}.  In contrast, no significant correlation between N/O and $\rm 12+log(O/H)$ was found for the sample of Seyfert nuclei from \citet{2017MNRAS.468L.113D}.  On the other hand, a negative correlation with $r=-0.45$ was derived for the LINERs in our sample, and with $r=-0.42$ when we combined the two samples compiled by  \cite{2021MNRAS.505.4289P} [see item (i) in Section \ref{n-o section}]. It worth to be mentioned that if only the sample of 45 LINERs at the local universe listed by \cite{2021MNRAS.505.4289P} is considered, no correlation is found.  The unexpected N/O negative trend was also reported by \citet{2018MNRAS.476.3793K} for the blue compact dwarf galaxy NGC\,4670 (and some other dwarf galaxies) as well as by \citet{2022AN....34320048Z} for the dwarf irregular galaxy NGC\,4163. Both galaxies were also studied performing a spatially-resolved analysis. It is important to emphasize that negative gradients found for these two dwarf galaxies were derived for objects showing significantly lower N/O and 12+log(O/H) values (which seems to be located close to the primary nitrogen production zone or in a transition zone where both mechanisms of nitrogen production could be of similar importance; see Fig.\ \ref{comparation}) than the results obtained for the LINERs studied in the present work.
  As mentioned by \citet{2018MNRAS.476.3793K}, the negative trend of N/O could be explained by various mechanisms, such as outflows,  star formation efficiency (SFE), and Wolf-Rayet (WR) stars, which are briefly discussed below. 

\begin{figure*}
\includegraphics*[width=.49\textwidth]{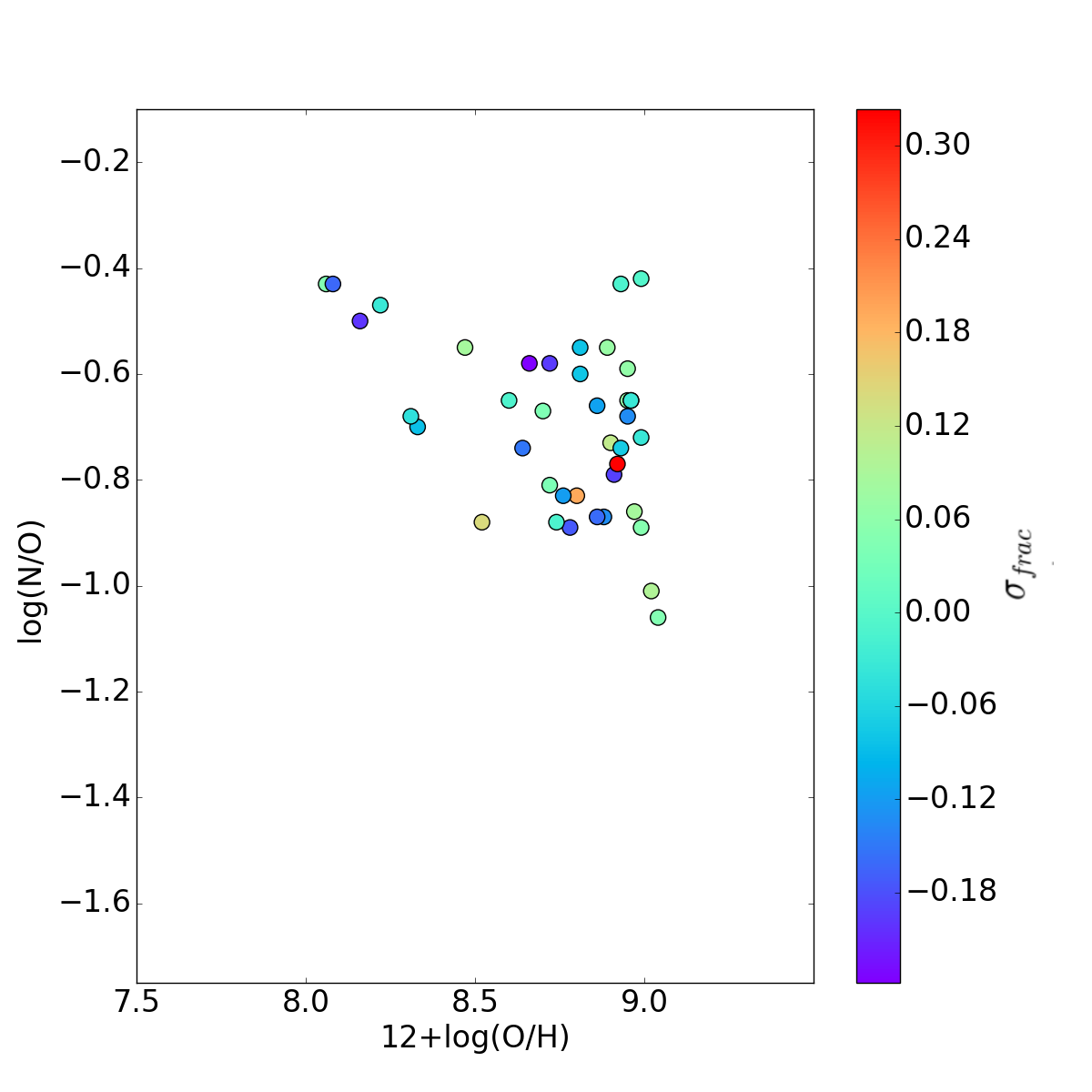}
\includegraphics*[width=.49\textwidth]{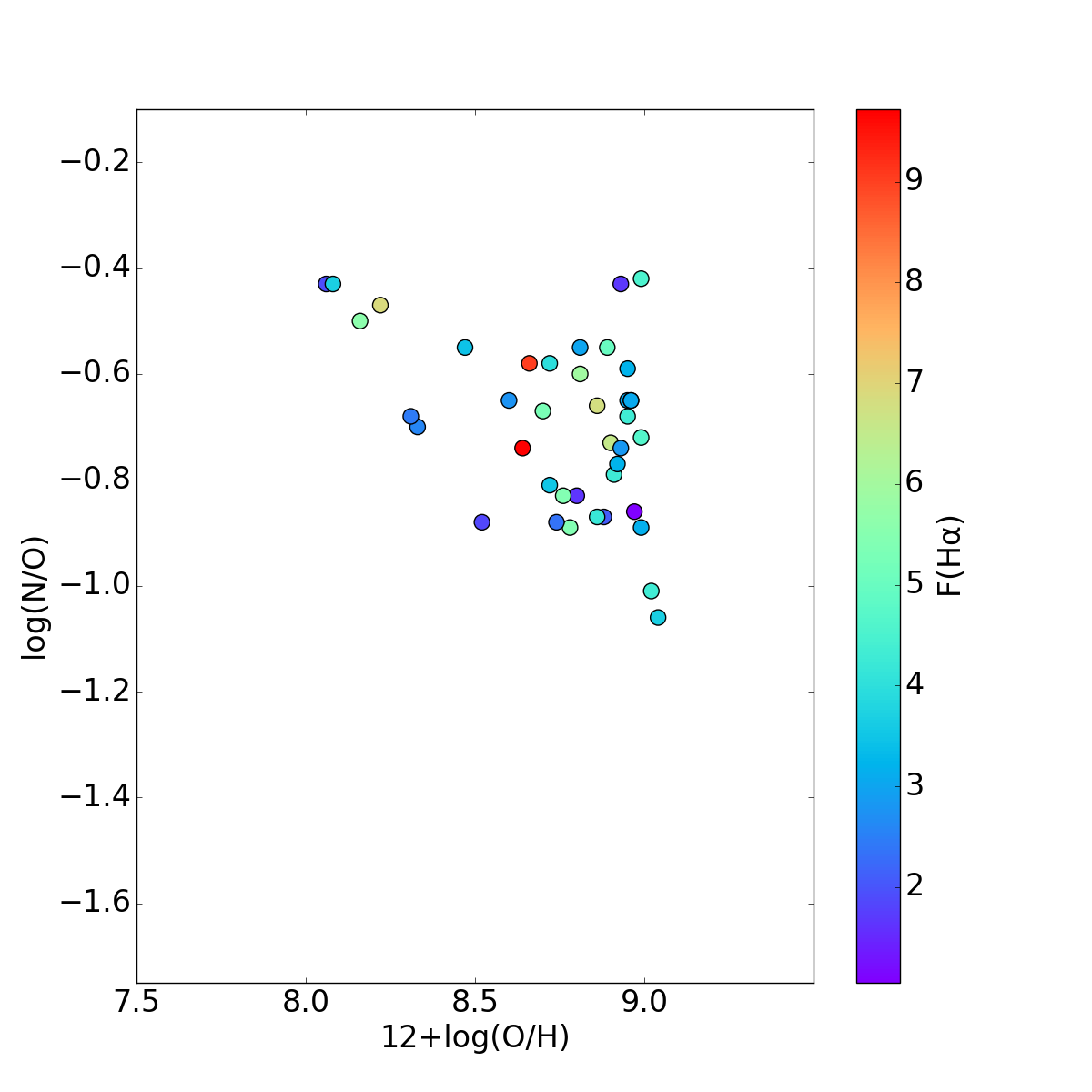}
\caption{log(N/O) vs.\ 12+log(O/H) diagram. The color bars indicate the $\rm \sigma_{frac} $ value (left panel) and the H$\alpha$ fluxes in units of 10$^{-17}$ erg/s/cm$^2$/spaxel (right panel) for each object in our sample. 
}
\label{sigma} 
\end{figure*}

To investigate the presence of outflows in our LINERs, we followed the approach by  \citet{2019MNRAS.484..252I}. These authors explored the impact of AGNs on the kinematics of their host galaxies, analyzing a sample with 62 AGNs and 109 inactive galaxies. They derived the fractional velocity dispersion differences between gas and stars [$\rm \sigma_{frac} = (\sigma_{gas}-\sigma_{stars})/\sigma_{stars}$]. A higher value of $\rm \sigma_{frac}$ indicates disturbed kinematics, likely due to outflows. \citet{2019MNRAS.484..252I}  found that 75\% of their AGNs have values of $\rm \sigma_{frac} > -0.13$, while 75\% of inactive galaxies have $\rm \sigma_{frac} < -0.04$. We computed  $\rm \sigma_{frac}$ for our sample of LINERs and found that 55\% of it has $\rm \sigma_{frac} > -0.04$. However, these higher values of $\rm \sigma_{frac} $ are not associated with higher values of N/O (see Fig.~\ref{sigma}, left panel) and we can not conclude that outflows are responsible for the negative trend on N/O.  About the SFE,  \citet{2018MNRAS.476.3793K} mentioned that, in the case of NGC\,4670, spaxels with higher N/O ratios and lower metallicities show higher H$\alpha$ fluxes, suggesting a connection between the N/O ratio and a very recent enhancement of the star formation rate. However, we did not find such a correlation for our LINERs (see right panel of Fig.~\ref{sigma}), likewise \citet{2022AN....34320048Z} did not find this correlation for NGC\,4163.  
Finally, regarding the presence of WR stars,  two important broad features in the optical spectra can reveal the presence of this kind of stars  \cite[see e.g.][]{2008A&A...491..131L}: the so-called blue WR bump (mainly formed by \ion{N}{v}, \ion{N}{iii}, \ion{C}{iii} + \ion{C}{iv} and \ion{He}{ii}$\lambda 4686$ emission lines) and the red WR bump (formed by the \ion{C}{v} emission lines). A visual inspection of our datacubes did not reveal any significant traces of these WR features. Therefore, the mechanisms highlighted by \citet{2018MNRAS.476.3793K} can not explain the negative trend in the N/O ratios observed in the present study.

 In a recent work, \citet{2023A&A...674L...7Z} found a significant scatter around the usual O/H-N/O curve in counter-rotating 
gas located in the central parts of galaxies. He suggested that such gas most likely fall onto these galaxies from areas outside their  disks. We searched for hints of co-rotation in the velocity maps of the LINERS in our sample and we did not find any evidence of this phenomenon.

\begin{figure}
\includegraphics*[width=.49\textwidth]{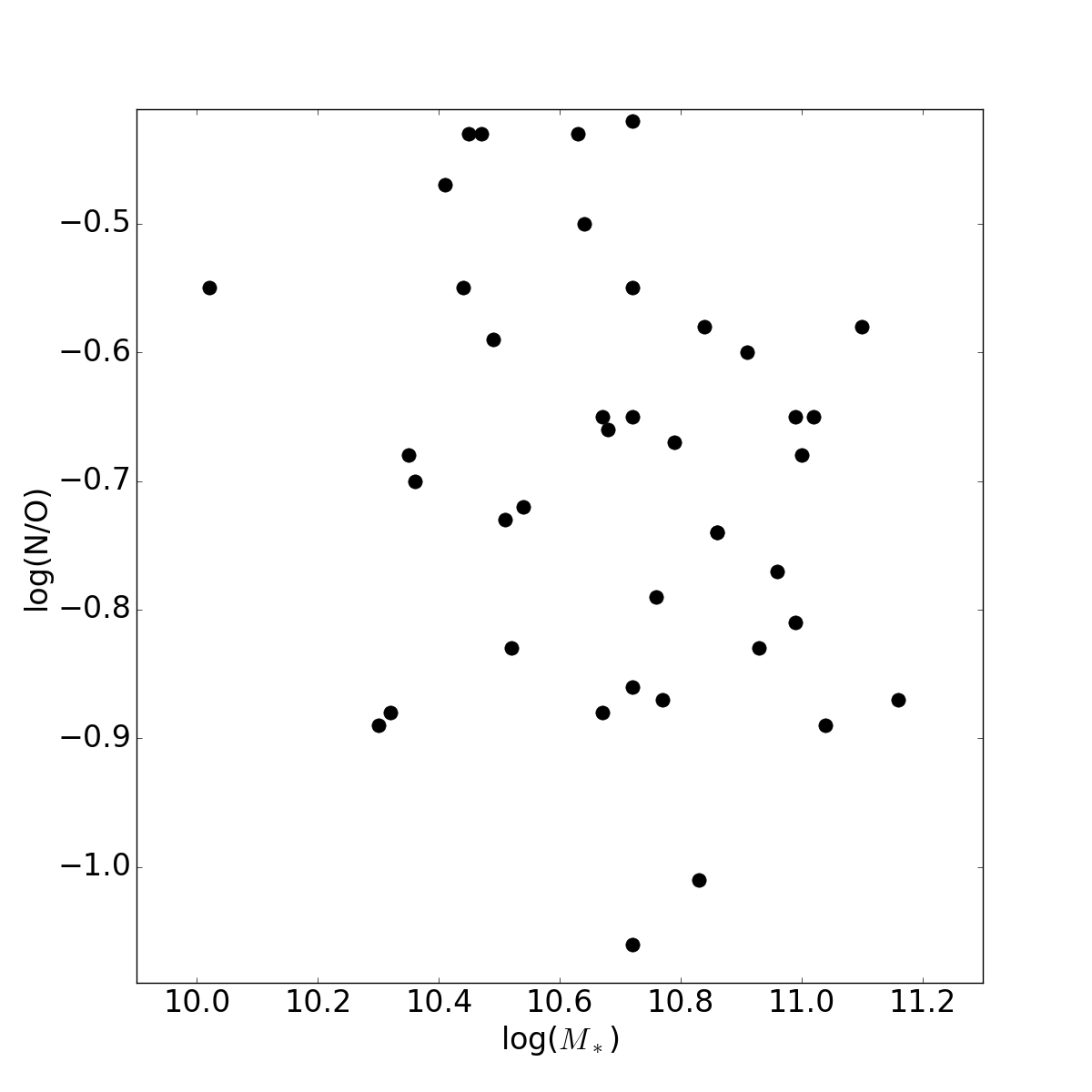}
\caption{log($M_{*}/\rm M_{\odot}$) vs.\ log(N/O). The mass corresponds to the stellar masses of the hosting galaxies of our sample and is taken from the manga.Pipe3D  and manga.drpall database \citep{2016RMxAA..52..171S}. The N/O abundance is taken from Table~\ref{saidas}.}
\label{mass} 
\end{figure}

We also investigated the correlation between the N/O abundances and the stellar masses of the host galaxies. The stellar masses of the hosting galaxies are in the range of $10\: \la \: \log(M_{*}/\rm M_{\odot})\: \la \: 11.2$ and were taken from the MaNGA   catalog Pipe3D\footnote{https://data.sdss.org/datamodel/files/MANGA\_PIPE3D} 
\citep{2016RMxAA..52..171S}. As is shown in Fig.~\ref{mass}, a large scatter was found and no correlation between these parameters is observed. A similar result was obtained by \citet{2021MNRAS.505.4289P} finding a very small  Pearson’s correlation coefficient ($r=0.038$) for their sample. It is important to note that our sample is composed only by massive galaxies, which occupy the flattened part of the Mass-Metallicity relation (MZR) curve, i.e., the region where no significant variation of metallicity ($\sim0.1$ dex) with the mass is found (see also \citealt{2019ApJ...874..100T}).  Likewise, as reported by \citet{2021MNRAS.505.4289P}, for SF galaxies there is a relation between N/O and stellar masses in the $7 \: \la \: \log(M_{*}/\rm M_{\odot}) \: \la \: 11$ range. On the other hand, these authors neither find any relation between N/O and stellar masses for AGN galaxies, i.e.,  Seyfert and LINER galaxies hosted in more massive galaxies ($9\: \la \: \log(M_{*}/\rm M_{\odot}) \: \la \:11)$.

\begin{table*}
\caption{Number of objects of each sample (N), median values and ranges for O/H and N/O abundances for each study:  SF galaxies reported by \citet{2021A&A...655A..58Z}; Seyfert 2 nuclei analyzed by \citet{2017MNRAS.468L.113D}, LINER galaxies studied by \citet{2021MNRAS.505.4289P}, and LINERs analyzed in this work. Fitted equations for each sample, being  $x = \rm 12+\log(O/H)$ and $ y=\rm \log(N/O)$, as well as Pearson's correlation coefficient, are also shown.}
\label{liners}
\begin{tabular}{@{}llc@{}cc@{}cc@{}cc@{}cccc@{}}
\hline
                             &      && \multicolumn{2}{@{}c@{}}{12+log(O/H)}       &&&          \multicolumn{2}{@{}c@{}}{log(N/O)}   & &           \multicolumn{1}{@{}c@{}}{Fitted equation}   &   \multicolumn{1}{c@{}}{Correlation Coeficient}\\
\cline{4-5}
\cline{8-9}
Reference                    & N    && Median            &     Range 	  &&&      Median         &      Range      &&  &  ($r$)  \\
\hline
\azul{Zinchenko et al. (2021)} & 1431 && $8.57\pm0.12$  &   $7.93,8.81$    &&&	$-0.86\pm0.22$     &  $-1.48,-0.24$  && $y=1.80x^2-28.91x+114.50$   & $0.92$ \\
\cite{2020ApJ...893...96B}   & 176  && $8.52\pm0.20$  &   $7.81,8.96$      &&&	$-0.92\pm0.24$     &  $-1.46,-0.42$ && $y=0.67x-6.61$              & $0.54$ \\
\cite{2017MNRAS.468L.113D}   & 44   && $8.65\pm0.19$  &	 $8.07,8.97$      &&&   $-0.54\pm0.12$     &  $-0.80,-0.29$ && $y=0.05x-0.97$              & $0.08$ \\
\cite{2021MNRAS.505.4289P}   & 65   && $8.57\pm0.24$  &	 $8.01,8.86$      &&&	$-0.61\pm0.20$     &  $-1.11,-0.21$ && $y=-0.34x+2.33$             & $-0.42$ \\
This work                    & 40   && $8.83\pm0.27$  &	 $8.05,9.03$      &&&	$-0.69\pm0.16$     &  $-1.05,-0.42$ && $y=-0.26x+1.66$             & $-0.45$ \\
\hline
\end{tabular}
\end{table*}

\section{Conclusion}
\label{conc}

Using optical data from 40 LINER galaxies taken from the MaNGA survey and classified as retired galaxies in the WHAN diagnostic diagram, we built detailed photoionization models using the {\sc cloudy} code to reproduce optical emission line intensities ratios of these objects. Based on these models we were able to estimate the oxygen and nitrogen abundances of the galaxies in our sample. We found 
that our LINER objects have oxygen and nitrogen abundances in the ranges of
$\rm 8.0 \: \la \: 12+\log(O/H) \: \la \: 9.0$ and $\rm 7.6 \: \la \: 12+\log(N/H) \: \la \: 8.5$, with mean values of $8.74\pm 0.27$ and $8.05\pm0.25$, respectively. We also found that about 70\% of the objects in our sample have oversolar oxygen and nitrogen abundances. The O/H and N/O abundance of our sample are in consonance with those obtained for Seyfert~2 nuclei as well as for the most metallic \ion{H}{ii} regions located in spiral galaxies. We
compared our results with the ones obtained for another LINERs sample whose estimates were taken from the literature. We found a very good agreement between each other.  The LINERs belonging to our sample are located in the higher N/O region of the O/H-N/O diagram presenting  a negative correlation between these two parameters with a correlation coefficient of $r=-0.45$. These results suggest that our objects mainly have a secondary nitrogen production together with some other mechanisms that deviate them from the usual theoretical secondary curve and from the  observational sequence of \ion{H}{ii} regions. We explore several mechanisms proposed in the literature to explain these deviations. We did not find any evidence to support the reported mechanisms.
We also investigated the existence of  N/O abundance dependence with the stellar masses of the hosting galaxies and we did not find any correlation between these two parameters.

\section*{Acknowledgements}
CBO is grateful to the Fundação de Amparo à Pesquisa do Estado de São Paulo (FAPESP) for the support under grant 2019/11934-0 and to the Coordenação de Aperfeiçoamento de Pessoal de Nível Superior (CAPES).  ACK thanks to FAPESP for the support grant 	2020/16416-5 and to Conselho Nacional de Desenvolvimento Científico e Tecnológico (CNPq). OLD is grateful to FAPESP and CNPq. JAHJ acknowledges support from FAPESP, process number 2021/08920-8. GSI acknowledges financial support from FAPESP (Proj. 2022/11799-9).

\section*{Data Availability}

The data underlying this article will be shared on reasonable request
to the corresponding author.




\bibliographystyle{mnras}
\bibliography{example} 



\bsp	
\label{lastpage}
\end{document}